\documentclass[twocolumn,english,aps,prl,floatfix,amssymb,superscriptaddress,longbibliography]{revtex4-2}
\usepackage[latin9]{inputenc}
\usepackage{verbatim}
\usepackage{float}
\usepackage{amsmath}
\usepackage{amssymb}
\usepackage{graphicx}
\usepackage{color}
\usepackage{xcolor}
\usepackage{tikz}
\usepackage[normalem]{ulem}

\newcommand{\ket}[1]{\ensuremath{\left| #1 \right>}}

\usepackage{bbold}

\usepackage[bookmarks=false,linkcolor=blue,urlcolor=blue,colorlinks,citecolor=blue]{hyperref}
\makeatletter

\newcommand{\be}{\begin{equation}}
\newcommand{\ee}{\end{equation}}
\newcommand{\bea}{\begin{eqnarray}}
\newcommand{\eea}{\end{eqnarray}}

\begin{document}

\title{Integrability of the Kondo model with time dependent interaction strength}
\author{Parameshwar R. Pasnoori}
\affiliation{Department of Physics, University of Maryland, College Park, MD 20742, United
States of America}
\affiliation{Laboratory for Physical Sciences, 8050 Greenmead Dr, College Park, MD 20740,
United States of America}
\email{pparmesh@umd.edu}
\begin{abstract}
In this letter we consider the time dependent Kondo model where a magnetic impurity interacts with the electrons through a time dependent interaction strength $J(t)$.  We develop a new framework based on Bethe ansatz and construct an exact solution to the time-dependent Schrodinger equation. We show that when periodic boundary conditions are applied, the consistency of the solution results in a constraint equation which relates the amplitudes corresponding to a certain ordering of the particles in the configuration space. This constraint equation takes the form of a matrix difference equation, and the associated consistency conditions restrict the interaction strength $J(t)$ for the system to be integrable. For a given $J(t)$ satisfying these constraints, the solution to the matrix difference equations provides the exact many-body wavefunction that satisfies the time-dependent Schrodinger equation. We provide a concrete example of $J(t)$ which satisfies these constraint equations. We show that in this case, the matrix difference equations turn into quantum Knizhnik-Zamolodchikov (qKZ) equations, which are well studied in the literature. The framework developed in this work allows one to probe the non-equilibrium physics of the Kondo model, and being general, it also allows one to solve new class of Hamiltonians with time-dependent interaction strength which are based on quantum Yang-Baxter algebra. 
\end{abstract}
\maketitle

\paragraph{Introduction}
Integrability plays an important role in understanding phenomena which are not amenable to regular analytical techniques. Integrable models provide insight into many-body effects and help in constructing and testing numerical techniques. There has been resurgence in studying integrable models \cite{Rylands_2024,TerrasXXZ1,Rylands20242,STSpaper,RylandsGalitski,parmeshkondo1,SSSTpaper,TerrasXXZ2,XXXmag,Kattel_2023,NHK2,kattel2024spin,Rylands_2022,parmeshkondo2,kattel2024,XXZpaper} due to the advancement in cold atom experiments \cite{coldatomsBA,Zhangcold,SPTcold} and quantum circuits which provide the possibility to engineer any model of interest \cite{KochpotSC,Quantumcircuitdynamics,LarkinSC,circuitSPT}. Bethe ansatz both in the coordinate and the algebraic form has played a pivotal role in constructing exact solutions to integrable models which are of interest to both the condensed matter and the high energy community. Arguably, the most famous examples being the Heisenberg spin chain \cite{Bethe1931}, Kondo model \cite{JunKondo} and the Thirring model \cite{THIRRING,pasnooriduality25} or equivalently the Sine-Gordon model \cite{ColemanSGMT}.  The quantum Yang-Baxter (QYB) equation \cite{Yang,BAXTER} is the backbone of any scattering process \cite{ZAMOLODCHIKOV} in an integrable model and plays a central role in the Bethe ansatz solution of these strongly interacting systems \cite{SklyaninQISM,Japaridze,Thacker}. In addition to the class of models based on QYB equation, there exists a different class of quantum integrable models based on the classical Yang-Baxter (CYB) equation, such as the Richardson-BCS model \cite{RICHARDSON}, Gaudin magnets etc \cite{Dukelsky,Gaudin1976}. 


In addition to phenomena which are described by Hamiltonians that are time-independent mentioned above, there exist processes which are effectively described by Hamiltonians with coupling strengths that are time-dependent. For example, in cold atoms experiments, the trapping potential can be time-dependent \cite{RAIZENcold} leading to a Hamiltonian with time-dependent interaction strength. In quantum circuits, the effective Hamiltonian describing the evolution of gates that are controlled and manipulated through rf pulsing is time-dependent \cite{circuitHamiltonian}. Hence, studying time-dependent Hamiltonians that are integrable is very crucial, as it allows one to obtain exact solutions of the time-dependent Schrodinger equation and allows one to probe non-equilibrium physics. All known time-dependent models that are integrable are based on the CYB equation, such as the time-dependent BCS and Dicke model \cite{YUZBASHYAN} and several other models which are of multi-level Landau Zener type \cite{demkov1968stationary,Patra_2015}. To the best of our knowledge, a solution to a model based on QYB equation whose interaction strength is time dependent is not known. 

\begin{center}
\begin{figure}
\includegraphics[width=1\columnwidth]{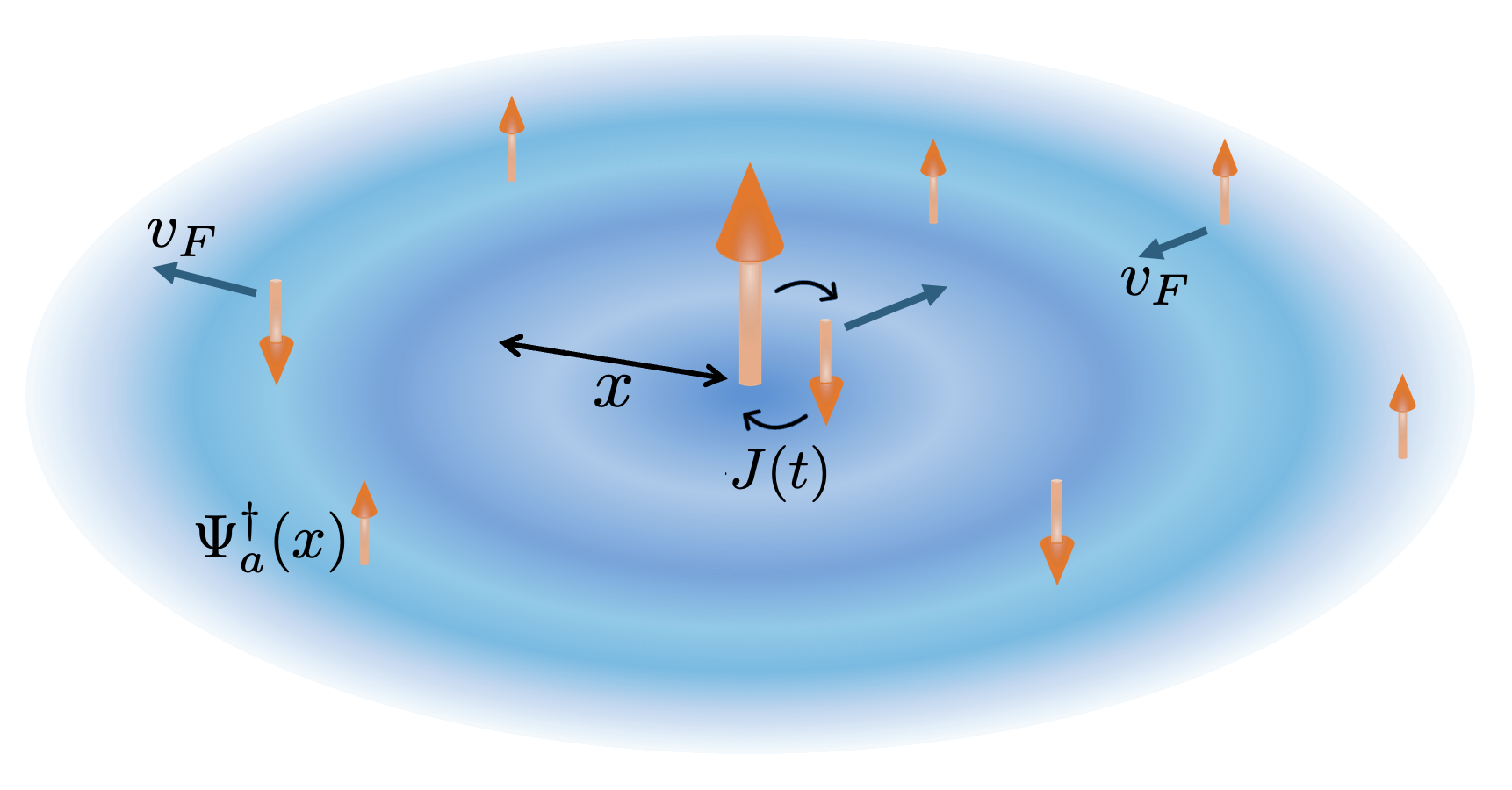}
\caption{Figure depicts electrons (small orange arrows) scattering off the magnetic impurity (big orange arrow). When an electron's position is close to the impurity, they interact through a spin-exchange interaction with time-dependent strength $J(t)$.  }
\label{fig:picture}
\end{figure}
\end{center}

In this letter we consider a model which falls in this category, namely the Kondo model with time dependent interaction strength. As the interaction strength is time-dependent, it allows one to study the non-equilibrium properties of the Kondo model. We develop a new frame work based on Bethe ansatz and construct an exact ansatz wavefunction which satisfies the time dependent Schrodinger equation. The wavefunction consists of several terms which correspond to different ordering of particles in the configuration space. Applying periodic boundary conditions and demanding the consistency of the solution results in a constraint equation between amplitudes corresponding to certain terms in the wavefunction, which takes the form of a matrix difference equation. These matrix difference equations give rise to certain consistency conditions that restrict the allowed functions for the interaction strength $J(t)$ such that the system is integrable. For a given $J(t)$ that satisfy these constraints, the matrix difference equation should be solved, which then yields the explicit form of the exact wavefunction that satisfies the time-dependent Schrodinger equation. Hence, the method developed in this work allows one to probe the non-equilibrium aspects of the Kondo model.

\paragraph{Hamiltonian}
Kondo model describes a magnetic impurity with spin $\frac{1}{2}$ interacting with a bath of non interacting conduction electrons through a spin exchange interaction (see Fig. (\ref{fig:picture})). At low temperature the Kondo effect takes place where the resistivity increases as the temperature is lowered and eventually saturates to a finite value at zero temperature. A strong coupling scale called the Kondo temperature is generated, below which, the impurity forms a many body singlet state with conductions electrons resulting in the screening of the impurity and increase in the resistivity \cite{Anderson,Wilson,Nozier}. The Kondo model has been solved exactly using the Bethe ansatz method \cite{Andrei80}\cite{Wiegmann_1981}. Here we consider this model with a time dependent interaction strength associated with the spin exchange interaction between the electrons and the impurity. The Hamiltonian is given by (\ref{Hamiltonian})
\bea \nonumber H=\int_{-(L-y)}^{y} \; dx\;  \big\{\Psi^{\dagger}_{a}(x)(-i\partial_x)\Psi_{a}(x) \\+ J(t) \Psi^{\dagger}_{a}(0)\left(\vec{\sigma}_{ab}\cdot\vec{S}_{\alpha\beta}\right)\Psi_{b}(0)\big\}, \label{Hamiltonian}
\eea
where $\Psi_{a}(x)$ describes the fermion (electron) field with subscript $a=\uparrow,\downarrow$ denoting the spin. $S$ represents the impurity and $J(t)$ is the time dependent interaction strength. For simplicity, we have set the Fermi velocity $v_F=1$. We have introduced the parameter $y$, where $0<y<L$ to keep the length of the system on either side of the impurity arbitrary. 

The system conserves the total number of electrons $N$
\bea N=\sum_{a}\int_{0}^{L} dx \Psi^{\dagger}_{a}(x)\Psi_{a}(x) 
\eea

and is spin rotation invariant as it commutes with the total spin operator $\vec S_{\rm T}= \vec s + \vec S_{}$, where $\vec S $ is the impurity spin operator and $\vec s_{} =   \int_{0}^L dx\; \vec s_{}(x)$ is that of the electrons, where
\be
\label{S}
\vec s_{}(x)= \frac{1}{2} (\Psi^{\dagger}_{a}(x) \vec \sigma_{ab} \Psi^{}_{b}(x) ).
\ee

Since the number of particles is a conserved quantity, we can look for wave function labeled by $N$ which satisfies the time dependent Schrodinger equation

\bea i\partial_t \ket{\Psi_N}= H \ket{\Psi_N},\label{SEk}\eea

where $H$ is the Hamiltonian (\ref{Hamiltonian}).

\paragraph{One particle solution and the S-matrix}

First consider the case of one particle. The wave function in the one particle sector can be written as
\bea \label{1pform}\ket{\Psi_1}=\sum_{a\alpha} \int_{-(L-y)}^y dx \; \Psi^{\dagger}_a(x) F_{a\alpha}(x,t)\ket{\alpha}.
\eea

Here $a,\alpha$ denote the spin degrees of freedom of the particle and the impurity respectively. Using the above expression in the Schrodinger equation (\ref{SEk}), we obtain 
\be-i(\partial_t+\partial_x)F_{a\alpha}(x,t)+J(t) \delta(x) \left(\vec{\sigma}_{ab}\cdot\vec{S}_{\alpha\beta}\right)F_{b\beta}(x,t)=0.\label{se1}\ee

To solve the above equation, we use the following ansatz (see supplementary material (SM) \cite{Supplemental} for detailed construction)
\begin{align}\nonumber F_{a\alpha}(x,t)=  \big(f^{10}_{a\alpha}(x-t)\theta(-x)\theta(L-y+x)\\ +f^{01}_{a\alpha}(x-t)\theta(x)\theta(y-x)\big).\label{1pformk}
\end{align}

Here $f^{10}_{a\alpha}(x-t),f^{01}_{a\alpha}(x-t)$ represent amplitudes corresponding to the particle being on the left and right sides of the impurity respectively. $\theta(x)$ is the Heaviside function where $\theta(x)=1$ for $x>0$ and $\theta(0)=1/2$. Note that the superscripts denote the ordering of the particle with respect to the impurity, where `1' represents the particle and `0' represents the impurity. Using the above expression (\ref{1pformk}) in the equation (\ref{se1}), we obtain the following relation between the two amplitudes in the one-particle wavefunction 

\bea \label{1prel}f^{01}_{a\alpha}(z)=S^{10}_{ab\alpha\beta}(z)f^{10}_{b\beta}(z),\eea
where we have used the notation $z=x-t$, and 
\begin{align} \label{smatk}&S^{10}(z)=e^{i\phi(z)}\frac{ig(z)I^{10}_{ab,\alpha\beta}+P^{10}_{ab,\alpha\beta}}{ig(z)+1},\\
&g(z)= \frac{1}{2J(t-x)}\left(1-\frac{3}{4}(J(t-x))^2\right),\\&e^{i\phi(z)}=\frac{1+\left(i/2J(t-x)\right)\left(1-(3/4)J^2(t-x)\right)}{iJ(t-x)-\left(1+(3/4)J^2(t-x)\right)}
\end{align}
is the particle-impurity S-matrix and the superscript denotes that it acts in the spin spaces of the particle `1' and the impurity `0'. In the above expression $I^{10}_{ab,\alpha\beta}$ is the identity operator and $P^{10}_{ab,\alpha\beta}$ is the permutation operator which acts in the spin spaces of the particle and the impurity and exchanges their spin. \footnote{ Explicit form of the permutation operator is
\be P^{ab}_{pq,rs}=\frac{1}{2}\left(I^{ab}_{pq,rs}+\sum_{\alpha=x,y,z}\sigma^{a,\alpha}_{pq}\sigma^{b,\alpha}_{rs}\right).
\ee
Here $\sigma^{a,\alpha}$ and $\sigma^{b,\alpha}$, $\alpha=x,y,z$ are the Pauli matrices acting in the spin spaces $a$ and $b$ respectively.} 

Hence, we find that the Hamiltonian relates the two amplitudes in the one particle wavefunction (\ref{1pformk}) through the particle-impurity S-matrix (\ref{smatk}), such that there exists one free amplitude. To obtain the explicit form of the one particle wavefunction one needs to determine this free amplitude. This can be achieved by applying periodic boundary conditions in the spatial direction, which yields the following constraint equation
\bea f^{10}_{a\alpha}(z-L)=f^{01}_{a\alpha}(z).\label{1pbck}
\eea
Using the above equation (\ref{1pbck}) in (\ref{1prel}), we arrive at the following relation
\bea \label{1pdiffk}f^{10}_{a\alpha}(z-L)=S^{10}_{ab\alpha\beta}(z)f^{10}_{b\beta}(z).\eea

In the special case where the interaction strength is constant $J(t)=J$, the functions $f^{10}_{a\alpha}(z),f^{01}_{a\alpha}(z)$ are simple exponentials
\bea f^{10}_{a\alpha}(z)=A^{10}_{a\alpha}e^{ikz},f^{01}_{a\alpha}(z)=A^{01}_{a\alpha}e^{ikz}.\eea
with $A^{01}_{a\alpha}=S^{10}_{ab\alpha\beta} A^{10}_{b\beta}$, where the S-matrix $S^{10}_{ab\alpha\beta}$ is of the same form as in (\ref{smatk}) with $J(t-x)=J$. In which case, (\ref{1pdiffk}) turns into an eigenvalue equation. In the general case where $J(t)$ is time dependent, the equation (\ref{1pdiffk}) is a matrix valued difference equation. Given a specific form of $J(t)$, one can solve (\ref{1pdiffk}) to obtain the amplitude $f^{10}_{a\alpha}(z)$. One can then use the relation (\ref{1prel}) to obtain the other amplitude and hence also the explicit form of the one particle wavefunction.

\paragraph{N particle solution and Yang-Baxter algebra}
Now let us consider the case of $N\ge 2$ number of particles. The wavefunction takes the form
\begin{align}\label{npform}\ket{\Psi_N}=\hspace{-0.15in}\sum_{\sigma_1...\sigma_N}\hspace{-0.8mm}\prod_{j=1}^{N}\int_{-(L-y)}^y\hspace{-0.3in}dx_j \Psi^{\dagger}_{\sigma_j}(x_j)\mathcal{A}F_{\sigma_1...\sigma_N\alpha}(x_1,...,x_N,t)\ket{\alpha}, 
\end{align}
where $\sigma_j$ denote the spin indices of the electrons and $\alpha$ denotes the spin of the impurity and $\mathcal{A}$ denotes anti-symmetrization with respect to $x_i$ and $\sigma_i$. Using the above expression (\ref{npform}) in the Schrodinger equation (\ref{SEk}), one obtains the following equation

\begin{align} &\nonumber-i(\partial_t+\sum_{j=1}^N\partial_{x_j})\mathcal{A}F_{\sigma_1...\sigma_N\alpha}(x_1,...,x_N,t)
\\\nonumber&+J(t)\big(\sum_{j=1}^N\delta(x_j)\:\vec{\sigma}_{\sigma_j\gamma_j}\cdot\vec{S}_{\alpha\beta}\:\prod_{l=1,l\neq j}^NI_{\sigma_l\gamma_l}\big)\\&\;\;\;\;\;\;\;\;\;\times \mathcal{A}F_{\gamma_1...\gamma_N\beta}(x_1,...,x_N,t)=0.\label{npdiffk}\end{align}
Unlike in the case of one particle, where we only had to distinguish between the amplitudes corresponding to the particle being on the left or right sides of the impurity, here one needs to distinguish between the amplitudes corresponding to different ordering of particles with respect to each other as well. We have 
\begin{align} 
F_{\sigma_1...\sigma_N}(x_1,...,x_N,t)
= \sum_Q \theta(\{x_{Q(j)}\})  f^Q_{\sigma_1...\sigma_N}(z_1,...,z_N).\label{npformexp}
\end{align}
where we used the notation $z_i=x_i-t$, $i=1,2$. In this expression, $Q$ denotes a permutation of the position orderings of particles and  $\theta(\{x_{Q(j)}\})$ is the Heaviside function that vanishes unless $x_{Q(1)} \le \dots \le x_{Q(N)}$. Here $f^{Q}_{\sigma_1...\sigma_N} (z_1,...,z_N)$ is the amplitude corresponding to the ordering of the particles denoted by $Q$.  Applying periodic boundary conditions in the spatial direction results in the following relations

\be f^{j...0}_{\sigma_1...\sigma_N}(z_1,...,z_j,..,z_N)=f^{...0j}_{\sigma_1...\sigma_N}(z_1,...,z_j+L,..,z_N).\label{pbcnp}\ee
 Here ``..." in the superscripts corresponds to any ordering of the rest of the particles, which is the same in the amplitudes on both sides of the equation. By using the expression (\ref{npformexp}) in the equation (\ref{npdiffk}), similar to the one particle case, we find that the amplitudes corresponding to a particle $j$ being on the left and right sides of the impurity are related through the particle-impurity S-matrix (\ref{smatk}) 
(for ease of notation, from here on we suppress the spin indices unless needed)


 \bea \label{nprel}f^{..0j..}(z)=S^{j0}(z)\:f^{..j0..}(z).\eea
 
 Here again ``.." in the superscripts corresponds to any ordering of the rest of the particles. As mentioned above, the consistency of the wave function required us to differentiate between the amplitudes which differ in the ordering of the particles with respect to each other. These pairs of amplitudes are not constrained by the Hamiltonian due to the relativistic dispersion \footnote{In the case of quadratic dispersion, the Hamiltonian fixes all the relations and this is one of the reasons why the regular Kondo model is not integrable with quadratic dispersion.}. To preserve integrability, one needs to choose a specific electron-electron S-matrix $S^{ij}(z_i,z_j)$
 that relates the amplitudes $f^{..ij..}(z_1,...,z_N)$ and $f^{..ji..}(z_1,...,z_N)$ which differ by the ordering of the two particles $i$ and $j$ with respect to each other 
 \begin{align}&
f^{..ji.}(z_1,...,z_N) =S^{ij}(z_i,z_j)f^{..ij..}(z_1,...,z_N),\label{eemateqsn}\end{align}
where
\begin{align}S^{ij}(z_i,z_j)=\frac{i\big(g(z_i)-g(z_j)\big)I^{ij}+P^{ij}}{ig(z_i)-ig(z_j)+1}.\label{smatee}
\end{align}
Here again ``.." in the superscripts corresponds to any ordering of the rest of the particles, which is the same in both amplitudes. The particle-impurity S-matrix (\ref{smatk}) and the particle-particle S-matrices (\ref{smatee}) satisfy the following Yang-Baxter algebra

\begin{align}S^{j0}(z_j)S^{i0}(z_i)S^{ij}(z_i,z_j)=S^{ij}(z_i,z_j)S^{i0}(z_i)S^{j0}(z_j)\label{YB2},\end{align}

\begin{align}\nonumber
&S^{ij}(z_i,z_j)S^{ik}(z_i,z_k)S^{jk}(z_j,z_k)
\\&=S^{jk}(z_j,z_k)S^{ik}(z_i,z_k)S^{ij}(z_i,z_j).\label{YB3}
\end{align}
Using the relations (\ref{nprel}) and (\ref{eemateqsn}), one can relate any amplitude in the N-particle wavefunction (\ref{npform}) in terms of one amplitude of our choice. Hence, similar to the one-particle case, we find that there exists one free amplitude. Without loss of generality, let us choose this amplitude to be $f^{N...10}_{\sigma_1,...,\sigma_N,\alpha}(x_1,...,x_N)$. Similar to the one particle case, this amplitude can be determined by applying periodic boundary conditions. Using the relation (\ref{pbcnp}), which is obtained by applying periodic boundary conditions, along with the relations (\ref{nprel}) and (\ref{eemateqsn}), one obtains the following equation
\begin{align} \nonumber &f^{N...j...10}_{\sigma_1...\sigma_N}(z_1,...,z_j-L,...,z_N)\\&=e^{i\phi(z_j)}Z_j(z_1,...,z_N) \; f^{N...j...10}_{\sigma_1...\sigma_N}(z_1,...,z_j,...,z_N),\label{GqKZ}\end{align}

where
\begin{widetext}
\begin{align}
    Z_j(z_1,...,z_N)=&\:e^{-i\phi(z_j)}\:    S^{jj+1}(z_j,z_{j+1}+L)S^{jj+2}(z_j,z_{j+2}+L)...S^{jN}(z_j,z_N+L)S^{j0}(z_j)S^{j1}(z_j,z_1)...S^{jj-1}(z_j,z_{j-1})
.\label{transfermat}\end{align}
\end{widetext}

The operator $e^{i\phi(z_j)}\:Z_{j}(z_1,...,z_N)$ transports the particle $j$ through the entire system once, when acting on the amplitude $f^{N...10}_{\sigma_1...\sigma_N}(z_1,...,z_N)$. 

\paragraph{Consistency conditions and constraints on integrability}
In order for the system of equations (\ref{GqKZ}) to be consistent and hence have a solution, the operators (\ref{transfermat}) should satisfy the following consistency conditions \footnote{These conditions impose that transporting particle `j' around the system first and then particle `i' around the system is equivalent to transporting particle `i' around the system first and then particle `j' around the system.}

\begin{align}\nonumber
&Z_i(z_1,...,z_j-L,...,z_N) Z_j(z_1,...,z_N)\\&=
Z_j(z_1,...,z_i-L,...,z_N) Z_i(z_1,...,z_N).\label{commutations}\end{align}

Using 
\begin{align} f^{N...10}(z_1,...,z_N)= A^{N...10}(z_1,...,z_N) \prod_{i=1}^N h(z_i)\end{align}
in the equation (\ref{GqKZ}), we have
\begin{align}\nonumber &A^{N...j...10}(z_1,...,z_j-L,...,z_N)\\&=Z_j(z_1,...,z_N) \: A^{N...j...10}(z_1,...,z_j,...,z_N),\label{diffeq1}\end{align}
and 
\begin{align}
h(z_j-L)=e^{i\phi (z_j)}h(z_j)\label{diffeq2}.\end{align}

Hence we find that the solution to the time dependent Schrodinger equation has resulted in two difference equations (\ref{diffeq1}) and (\ref{diffeq2}), which correspond to the spin and the phase part respectively. Note that in the case where the interaction strength in the Hamiltonian is time independent, the equation (\ref{diffeq1}) reduces to an eigenvalue equation and the operator (\ref{transfermat}) is identified with the monodromy matrix \cite{Andrei80}. In which case, the arguments of the S-matrices (\ref{smatk}) and (\ref{smatee}) are constants, and the Yang-Baxters equations (\ref{YB2}) and (\ref{YB3}), are sufficient to show that the transfer matrices with different spectral parameters commute, which guarantees that the system is integrable.    

In the current case of time-dependent interaction strength, one needs to solve the equations (\ref{diffeq1}) and (\ref{diffeq2}). Equation (\ref{diffeq2}) which is the phase part of equation (\ref{GqKZ}), is an analytic difference equation and has been well studied for different classes of functions $\phi(x)$ \cite{ruijisenars}. To solve equation (\ref{diffeq1}), the existence of the Yang-Baxters equations (\ref{YB2}) and (\ref{YB3}) is not sufficient. This should be expected since they do not impose any restriction on $J(t)$. In order for the system of equations (\ref{diffeq1}) to be consistent and hence have a solution, the operators (\ref{transfermat}) should satisfy the consistency conditions (\ref{commutations}).
These conditions are satisfied by (\ref{transfermat}) only for specific functions of $J(t)$.  Hence, the conditions (\ref{commutations}) can be considered as the constraints on the interaction strength $J(t)$ for the system to be integrable. One such example is \footnote{The case where $a,c \in \mathbb{C}$ is also a solution, in which case, the Hamiltonian is non Hermitian and is interesting on its own right.}  
(See SM \cite{Supplemental} for more details)
\be J(t)=\frac{c}{a+t}, \:\:a,c \in \mathbb{R},\: c\ll1, \:a>0. \label{example}\ee 
 In this case the difference equations (\ref{diffeq1}) reduce to \textit{quantum Knizhnik-Zamolodchikov (qKZ) equations} \cite{rishetikhin1}. The resulting qKZ equations and the analytic difference equations (\ref{diffeq2}) should be solved to obtain the amplitude $f^{N,...j,...10}_{\sigma_1...\sigma_N}(z_1,..,z_j,..,z_N)$ in (\ref{GqKZ}). One can then use the relations (\ref{nprel}) and (\ref{eemateqsn}) to obtain the rest of the amplitudes in the N-particle wavefunction (\ref{npform}), and thereby obtain the exact many-body wavefunction that satisfies the time-dependent Schrodinger equation.  The qKZ equations first appeared in \cite{Smirnov_1986} as the fundamental equations for form factors in the sine-Gordon model, and were later derived from representation theory of quantum affine algebras \cite{Frenkel}. They have been well studied in the literature \cite{rishetikhin1,rishetikhin2,varchenko,matsuo} and the off-shell Bethe ansatz method to solve them has been developed in \cite{Babujian_1997}. In the forthcoming work \cite{ParmeshEmil}, we study the most general integrable interaction strengths $J(t)$, and solve the qKZ equations using the off-shell Bethe ansatz method and obtain the explicit form of the exact many body wavefunction which satisfies the time-dependent Schrodinger equation.

\paragraph{Discussion}
In this work we have considered the Kondo model with time-dependent interaction strength. We developed a framework based on the Bethe ansatz using which we obtained the exact ansatz wavefunction which satisfies the time-dependent Schrodinger equation. Applying periodic boundary conditions and demanding the consistency of the wavefunction results in a constraint equation between certain amplitudes in the wavefunction which takes the form of matrix difference equations. These matrix difference equations give rise to certain consistency conditions which restrict the allowed functions of the interaction strength $J(t)$ for the system to be integrable. We showed that for certain $J(t)$ satisfying these integrability constraints, the matrix difference equations turn into quantum Knizhnik-Zamolodchikov (qKZ) equations. These qKZ equations can be solved by the off-shell Bethe ansatz method, thus yielding the explicit form of many-body wavefunction that satisfies the time-dependent Schrodinger equation. 

Our work provides a general framework to probe the non-equilibrium aspects of the Kondo model, and also provides a new method to
solve a new class of Hamiltonians with time-dependent interaction strength that are based on quantum Yang-Baxter algebra such as the $SU(N)$ Gross-Neveu model and the sine-Gordon model. These paradigmatic models exhibit very rich phenomena, such as symmetry protected topological (SPT) phases \cite{SSSTpaper,pasnooriduality25} when the interaction strengths are constant. This naturally leads to the following question: Do these models exhibit a new type of interesting phase similar to SPT phase in the presence of time-dependent interaction strength, and if so, how do the associated edge modes and the entanglement structure depend on time. Another possible application is the spin chains with time-dependent interaction strength, such as the spin $1/2$ XXZ chain. It is well known that this system exhibits a spontaneous symmetry breaking of the discrete $Z_2$ spin-flip symmetry in the gapped regime, where it exhibits strong zero modes \cite{Fendley} and spin fractionalization with spin $1/4$ at the boundaries \cite{XXZpaper}. It would be interesting to study the stability of the symmetry broken phase in the presence of time-dependent interaction strength and its effect on the spin fractionalization and the strong zero modes.

\acknowledgments
I acknowledge very helpful and enlightening discussions with P. Azaria, C. Rylands and E. A. Yuzbashyan. I gratefully acknowledge support from the Simons Center for Geometry and Physics, Stony Brook University while attending the workshop `New Directions in far from Equilibrium Integrability and beyond' at which some of the research for this paper was performed.
\bibliography{refpaperTDK}

\begin{thebibliography}{62}%
\makeatletter
\providecommand \@ifxundefined [1]{%
 \@ifx{#1\undefined}
}%
\providecommand \@ifnum [1]{%
 \ifnum #1\expandafter \@firstoftwo
 \else \expandafter \@secondoftwo
 \fi
}%
\providecommand \@ifx [1]{%
 \ifx #1\expandafter \@firstoftwo
 \else \expandafter \@secondoftwo
 \fi
}%
\providecommand \natexlab [1]{#1}%
\providecommand \enquote  [1]{``#1''}%
\providecommand \bibnamefont  [1]{#1}%
\providecommand \bibfnamefont [1]{#1}%
\providecommand \citenamefont [1]{#1}%
\providecommand \href@noop [0]{\@secondoftwo}%
\providecommand \href [0]{\begingroup \@sanitize@url \@href}%
\providecommand \@href[1]{\@@startlink{#1}\@@href}%
\providecommand \@@href[1]{\endgroup#1\@@endlink}%
\providecommand \@sanitize@url [0]{\catcode `\\12\catcode `\$12\catcode
  `\&12\catcode `\#12\catcode `\^12\catcode `\_12\catcode `\%12\relax}%
\providecommand \@@startlink[1]{}%
\providecommand \@@endlink[0]{}%
\providecommand \url  [0]{\begingroup\@sanitize@url \@url }%
\providecommand \@url [1]{\endgroup\@href {#1}{\urlprefix }}%
\providecommand \urlprefix  [0]{URL }%
\providecommand \Eprint [0]{\href }%
\providecommand \doibase [0]{https://doi.org/}%
\providecommand \selectlanguage [0]{\@gobble}%
\providecommand \bibinfo  [0]{\@secondoftwo}%
\providecommand \bibfield  [0]{\@secondoftwo}%
\providecommand \translation [1]{[#1]}%
\providecommand \BibitemOpen [0]{}%
\providecommand \bibitemStop [0]{}%
\providecommand \bibitemNoStop [0]{.\EOS\space}%
\providecommand \EOS [0]{\spacefactor3000\relax}%
\providecommand \BibitemShut  [1]{\csname bibitem#1\endcsname}%
\let\auto@bib@innerbib\@empty
\bibitem [{\citenamefont {Rylands}\ \emph
  {et~al.}(2024{\natexlab{a}})\citenamefont {Rylands}, \citenamefont
  {Vernier},\ and\ \citenamefont {Calabrese}}]{Rylands_2024}%
  \BibitemOpen
  \bibfield  {author} {\bibinfo {author} {\bibfnamefont {C.}~\bibnamefont
  {Rylands}}, \bibinfo {author} {\bibfnamefont {E.}~\bibnamefont {Vernier}},\
  and\ \bibinfo {author} {\bibfnamefont {P.}~\bibnamefont {Calabrese}},\
  }\bibfield  {title} {\bibinfo {title} {Dynamical symmetry restoration in the
  heisenberg spin chain},\ }\href {https://doi.org/10.1088/1742-5468/ad97b3}
  {\bibfield  {journal} {\bibinfo  {journal} {Journal of Statistical Mechanics:
  Theory and Experiment}\ }\textbf {\bibinfo {volume} {2024}},\ \bibinfo
  {pages} {123102} (\bibinfo {year} {2024}{\natexlab{a}})}\BibitemShut
  {NoStop}%
\bibitem [{\citenamefont {Grijalva}\ \emph {et~al.}(2019)\citenamefont
  {Grijalva}, \citenamefont {Nardis},\ and\ \citenamefont
  {Terras}}]{TerrasXXZ1}%
  \BibitemOpen
  \bibfield  {author} {\bibinfo {author} {\bibfnamefont {S.}~\bibnamefont
  {Grijalva}}, \bibinfo {author} {\bibfnamefont {J.~D.}\ \bibnamefont
  {Nardis}},\ and\ \bibinfo {author} {\bibfnamefont {V.}~\bibnamefont
  {Terras}},\ }\bibfield  {title} {\bibinfo {title} {{Open XXZ chain and
  boundary modes at zero temperature}},\ }\href
  {https://doi.org/10.21468/SciPostPhys.7.2.023} {\bibfield  {journal}
  {\bibinfo  {journal} {SciPost Phys.}\ }\textbf {\bibinfo {volume} {7}},\
  \bibinfo {pages} {023} (\bibinfo {year} {2019})}\BibitemShut {NoStop}%
\bibitem [{\citenamefont {Rylands}\ \emph
  {et~al.}(2024{\natexlab{b}})\citenamefont {Rylands}, \citenamefont {Klobas},
  \citenamefont {Ares}, \citenamefont {Calabrese}, \citenamefont {Murciano},\
  and\ \citenamefont {Bertini}}]{Rylands20242}%
  \BibitemOpen
  \bibfield  {author} {\bibinfo {author} {\bibfnamefont {C.}~\bibnamefont
  {Rylands}}, \bibinfo {author} {\bibfnamefont {K.}~\bibnamefont {Klobas}},
  \bibinfo {author} {\bibfnamefont {F.}~\bibnamefont {Ares}}, \bibinfo {author}
  {\bibfnamefont {P.}~\bibnamefont {Calabrese}}, \bibinfo {author}
  {\bibfnamefont {S.}~\bibnamefont {Murciano}},\ and\ \bibinfo {author}
  {\bibfnamefont {B.}~\bibnamefont {Bertini}},\ }\bibfield  {title} {\bibinfo
  {title} {Microscopic origin of the quantum mpemba effect in integrable
  systems},\ }\href {https://doi.org/10.1103/PhysRevLett.133.010401} {\bibfield
   {journal} {\bibinfo  {journal} {Phys. Rev. Lett.}\ }\textbf {\bibinfo
  {volume} {133}},\ \bibinfo {pages} {010401} (\bibinfo {year}
  {2024}{\natexlab{b}})}\BibitemShut {NoStop}%
\bibitem [{\citenamefont {Pasnoori}\ \emph
  {et~al.}(2020{\natexlab{a}})\citenamefont {Pasnoori}, \citenamefont
  {Andrei},\ and\ \citenamefont {Azaria}}]{STSpaper}%
  \BibitemOpen
  \bibfield  {author} {\bibinfo {author} {\bibfnamefont {P.~R.}\ \bibnamefont
  {Pasnoori}}, \bibinfo {author} {\bibfnamefont {N.}~\bibnamefont {Andrei}},\
  and\ \bibinfo {author} {\bibfnamefont {P.}~\bibnamefont {Azaria}},\
  }\bibfield  {title} {\bibinfo {title} {Edge modes in one-dimensional
  topological charge conserving spin-triplet superconductors: Exact results
  from bethe ansatz},\ }\href {https://doi.org/10.1103/PhysRevB.102.214511}
  {\bibfield  {journal} {\bibinfo  {journal} {Phys. Rev. B}\ }\textbf {\bibinfo
  {volume} {102}},\ \bibinfo {pages} {214511} (\bibinfo {year}
  {2020}{\natexlab{a}})}\BibitemShut {NoStop}%
\bibitem [{\citenamefont {Parhizkar}\ \emph {et~al.}(2024)\citenamefont
  {Parhizkar}, \citenamefont {Rylands},\ and\ \citenamefont
  {Galitski}}]{RylandsGalitski}%
  \BibitemOpen
  \bibfield  {author} {\bibinfo {author} {\bibfnamefont {A.}~\bibnamefont
  {Parhizkar}}, \bibinfo {author} {\bibfnamefont {C.}~\bibnamefont {Rylands}},\
  and\ \bibinfo {author} {\bibfnamefont {V.}~\bibnamefont {Galitski}},\
  }\bibfield  {title} {\bibinfo {title} {Path integral approach to quantum
  anomalies in interacting models},\ }\href
  {https://doi.org/10.1103/PhysRevB.109.155109} {\bibfield  {journal} {\bibinfo
   {journal} {Phys. Rev. B}\ }\textbf {\bibinfo {volume} {109}},\ \bibinfo
  {pages} {155109} (\bibinfo {year} {2024})}\BibitemShut {NoStop}%
\bibitem [{\citenamefont {Pasnoori}\ \emph
  {et~al.}(2020{\natexlab{b}})\citenamefont {Pasnoori}, \citenamefont
  {Rylands},\ and\ \citenamefont {Andrei}}]{parmeshkondo1}%
  \BibitemOpen
  \bibfield  {author} {\bibinfo {author} {\bibfnamefont {P.~R.}\ \bibnamefont
  {Pasnoori}}, \bibinfo {author} {\bibfnamefont {C.}~\bibnamefont {Rylands}},\
  and\ \bibinfo {author} {\bibfnamefont {N.}~\bibnamefont {Andrei}},\
  }\bibfield  {title} {\bibinfo {title} {Kondo impurity at the edge of a
  superconducting wire},\ }\href
  {https://doi.org/10.1103/PhysRevResearch.2.013006} {\bibfield  {journal}
  {\bibinfo  {journal} {Phys. Rev. Res.}\ }\textbf {\bibinfo {volume} {2}},\
  \bibinfo {pages} {013006} (\bibinfo {year} {2020}{\natexlab{b}})}\BibitemShut
  {NoStop}%
\bibitem [{\citenamefont {Pasnoori}\ \emph {et~al.}(2021)\citenamefont
  {Pasnoori}, \citenamefont {Andrei},\ and\ \citenamefont
  {Azaria}}]{SSSTpaper}%
  \BibitemOpen
  \bibfield  {author} {\bibinfo {author} {\bibfnamefont {P.~R.}\ \bibnamefont
  {Pasnoori}}, \bibinfo {author} {\bibfnamefont {N.}~\bibnamefont {Andrei}},\
  and\ \bibinfo {author} {\bibfnamefont {P.}~\bibnamefont {Azaria}},\
  }\bibfield  {title} {\bibinfo {title} {Boundary-induced topological and
  mid-gap states in charge conserving one-dimensional superconductors:
  Fractionalization transition},\ }\href
  {https://doi.org/10.1103/PhysRevB.104.134519} {\bibfield  {journal} {\bibinfo
   {journal} {Phys. Rev. B}\ }\textbf {\bibinfo {volume} {104}},\ \bibinfo
  {pages} {134519} (\bibinfo {year} {2021})}\BibitemShut {NoStop}%
\bibitem [{\citenamefont {Abetian}\ \emph {et~al.}(2025)\citenamefont
  {Abetian}, \citenamefont {Kitanine},\ and\ \citenamefont
  {Terras}}]{TerrasXXZ2}%
  \BibitemOpen
  \bibfield  {author} {\bibinfo {author} {\bibfnamefont {C.}~\bibnamefont
  {Abetian}}, \bibinfo {author} {\bibfnamefont {N.}~\bibnamefont {Kitanine}},\
  and\ \bibinfo {author} {\bibfnamefont {V.}~\bibnamefont {Terras}},\
  }\bibfield  {title} {\bibinfo {title} {{Boundary overlap in the open XXZ spin
  chain}},\ }\href {https://doi.org/10.21468/SciPostPhys.18.1.026} {\bibfield
  {journal} {\bibinfo  {journal} {SciPost Phys.}\ }\textbf {\bibinfo {volume}
  {18}},\ \bibinfo {pages} {026} (\bibinfo {year} {2025})}\BibitemShut
  {NoStop}%
\bibitem [{\citenamefont {Pasnoori}\ \emph {et~al.}(2023)\citenamefont
  {Pasnoori}, \citenamefont {Lee}, \citenamefont {Pixley}, \citenamefont
  {Andrei},\ and\ \citenamefont {Azaria}}]{XXXmag}%
  \BibitemOpen
  \bibfield  {author} {\bibinfo {author} {\bibfnamefont {P.~R.}\ \bibnamefont
  {Pasnoori}}, \bibinfo {author} {\bibfnamefont {J.}~\bibnamefont {Lee}},
  \bibinfo {author} {\bibfnamefont {J.~H.}\ \bibnamefont {Pixley}}, \bibinfo
  {author} {\bibfnamefont {N.}~\bibnamefont {Andrei}},\ and\ \bibinfo {author}
  {\bibfnamefont {P.}~\bibnamefont {Azaria}},\ }\bibfield  {title} {\bibinfo
  {title} {Boundary quantum phase transitions in the spin-$\frac{1}{2}$
  heisenberg chain with boundary magnetic fields},\ }\href
  {https://doi.org/10.1103/PhysRevB.107.224412} {\bibfield  {journal} {\bibinfo
   {journal} {Phys. Rev. B}\ }\textbf {\bibinfo {volume} {107}},\ \bibinfo
  {pages} {224412} (\bibinfo {year} {2023})}\BibitemShut {NoStop}%
\bibitem [{\citenamefont {Kattel}\ \emph {et~al.}(2023)\citenamefont {Kattel},
  \citenamefont {Pasnoori},\ and\ \citenamefont {Andrei}}]{Kattel_2023}%
  \BibitemOpen
  \bibfield  {author} {\bibinfo {author} {\bibfnamefont {P.}~\bibnamefont
  {Kattel}}, \bibinfo {author} {\bibfnamefont {P.~R.}\ \bibnamefont
  {Pasnoori}},\ and\ \bibinfo {author} {\bibfnamefont {N.}~\bibnamefont
  {Andrei}},\ }\bibfield  {title} {\bibinfo {title} {Exact solution of a
  non-hermitian pt-symmetric spin chain},\ }\href
  {https://doi.org/10.1088/1751-8121/ace56e} {\bibfield  {journal} {\bibinfo
  {journal} {Journal of Physics A: Mathematical and Theoretical}\ }\textbf
  {\bibinfo {volume} {56}},\ \bibinfo {pages} {325001} (\bibinfo {year}
  {2023})}\BibitemShut {NoStop}%
\bibitem [{\citenamefont {Kattel}\ \emph
  {et~al.}(2025{\natexlab{a}})\citenamefont {Kattel}, \citenamefont {Zhakenov},
  \citenamefont {Pasnoori}, \citenamefont {Azaria},\ and\ \citenamefont
  {Andrei}}]{NHK2}%
  \BibitemOpen
  \bibfield  {author} {\bibinfo {author} {\bibfnamefont {P.}~\bibnamefont
  {Kattel}}, \bibinfo {author} {\bibfnamefont {A.}~\bibnamefont {Zhakenov}},
  \bibinfo {author} {\bibfnamefont {P.~R.}\ \bibnamefont {Pasnoori}}, \bibinfo
  {author} {\bibfnamefont {P.}~\bibnamefont {Azaria}},\ and\ \bibinfo {author}
  {\bibfnamefont {N.}~\bibnamefont {Andrei}},\ }\bibfield  {title} {\bibinfo
  {title} {Dissipation driven phase transition in the non-hermitian kondo
  model},\ }\href {https://doi.org/10.1103/PhysRevB.111.L201106} {\bibfield
  {journal} {\bibinfo  {journal} {Phys. Rev. B}\ }\textbf {\bibinfo {volume}
  {111}},\ \bibinfo {pages} {L201106} (\bibinfo {year}
  {2025}{\natexlab{a}})}\BibitemShut {NoStop}%
\bibitem [{\citenamefont {Kattel}\ \emph
  {et~al.}(2025{\natexlab{b}})\citenamefont {Kattel}, \citenamefont {Pasnoori},
  \citenamefont {Pixley},\ and\ \citenamefont {Andrei}}]{kattel2024spin}%
  \BibitemOpen
  \bibfield  {author} {\bibinfo {author} {\bibfnamefont {P.}~\bibnamefont
  {Kattel}}, \bibinfo {author} {\bibfnamefont {P.~R.}\ \bibnamefont
  {Pasnoori}}, \bibinfo {author} {\bibfnamefont {J.~H.}\ \bibnamefont
  {Pixley}},\ and\ \bibinfo {author} {\bibfnamefont {N.}~\bibnamefont
  {Andrei}},\ }\bibfield  {title} {\bibinfo {title} {Spin chain with
  non-hermitian $\mathcal{PT}$-symmetric boundary couplings: Exact solution,
  dissipative kondo effect, and phase transitions on the edge},\ }\href
  {https://doi.org/10.1103/PhysRevB.111.224407} {\bibfield  {journal} {\bibinfo
   {journal} {Phys. Rev. B}\ }\textbf {\bibinfo {volume} {111}},\ \bibinfo
  {pages} {224407} (\bibinfo {year} {2025}{\natexlab{b}})}\BibitemShut
  {NoStop}%
\bibitem [{\citenamefont {Rylands}\ \emph {et~al.}(2022)\citenamefont
  {Rylands}, \citenamefont {Bertini},\ and\ \citenamefont
  {Calabrese}}]{Rylands_2022}%
  \BibitemOpen
  \bibfield  {author} {\bibinfo {author} {\bibfnamefont {C.}~\bibnamefont
  {Rylands}}, \bibinfo {author} {\bibfnamefont {B.}~\bibnamefont {Bertini}},\
  and\ \bibinfo {author} {\bibfnamefont {P.}~\bibnamefont {Calabrese}},\
  }\bibfield  {title} {\bibinfo {title} {Integrable quenches in the hubbard
  model},\ }\href {https://doi.org/10.1088/1742-5468/ac98be} {\bibfield
  {journal} {\bibinfo  {journal} {Journal of Statistical Mechanics: Theory and
  Experiment}\ }\textbf {\bibinfo {volume} {2022}},\ \bibinfo {pages} {103103}
  (\bibinfo {year} {2022})}\BibitemShut {NoStop}%
\bibitem [{\citenamefont {Pasnoori}\ \emph {et~al.}(2022)\citenamefont
  {Pasnoori}, \citenamefont {Andrei}, \citenamefont {Rylands},\ and\
  \citenamefont {Azaria}}]{parmeshkondo2}%
  \BibitemOpen
  \bibfield  {author} {\bibinfo {author} {\bibfnamefont {P.~R.}\ \bibnamefont
  {Pasnoori}}, \bibinfo {author} {\bibfnamefont {N.}~\bibnamefont {Andrei}},
  \bibinfo {author} {\bibfnamefont {C.}~\bibnamefont {Rylands}},\ and\ \bibinfo
  {author} {\bibfnamefont {P.}~\bibnamefont {Azaria}},\ }\bibfield  {title}
  {\bibinfo {title} {Rise and fall of yu-shiba-rusinov bound states in
  charge-conserving s-wave one-dimensional superconductors},\ }\href
  {https://doi.org/10.1103/PhysRevB.105.174517} {\bibfield  {journal} {\bibinfo
   {journal} {Phys. Rev. B}\ }\textbf {\bibinfo {volume} {105}},\ \bibinfo
  {pages} {174517} (\bibinfo {year} {2022})}\BibitemShut {NoStop}%
\bibitem [{\citenamefont {Kattel}\ \emph
  {et~al.}(2025{\natexlab{c}})\citenamefont {Kattel}, \citenamefont {Pasnoori},
  \citenamefont {Pixley},\ and\ \citenamefont {Andrei}}]{kattel2024}%
  \BibitemOpen
  \bibfield  {author} {\bibinfo {author} {\bibfnamefont {P.}~\bibnamefont
  {Kattel}}, \bibinfo {author} {\bibfnamefont {P.~R.}\ \bibnamefont
  {Pasnoori}}, \bibinfo {author} {\bibfnamefont {J.~H.}\ \bibnamefont
  {Pixley}},\ and\ \bibinfo {author} {\bibfnamefont {N.}~\bibnamefont
  {Andrei}},\ }\bibfield  {title} {\bibinfo {title} {Edge modes and boundary
  impurities in the anisotropic heisenberg spin chain},\ }\href
  {https://doi.org/10.1103/PhysRevB.111.174430} {\bibfield  {journal} {\bibinfo
   {journal} {Phys. Rev. B}\ }\textbf {\bibinfo {volume} {111}},\ \bibinfo
  {pages} {174430} (\bibinfo {year} {2025}{\natexlab{c}})}\BibitemShut
  {NoStop}%
\bibitem [{\citenamefont {Pasnoori}\ \emph
  {et~al.}(2025{\natexlab{a}})\citenamefont {Pasnoori}, \citenamefont {Tang},
  \citenamefont {Lee}, \citenamefont {Pixley}, \citenamefont {Andrei},\ and\
  \citenamefont {Azaria}}]{XXZpaper}%
  \BibitemOpen
  \bibfield  {author} {\bibinfo {author} {\bibfnamefont {P.~R.}\ \bibnamefont
  {Pasnoori}}, \bibinfo {author} {\bibfnamefont {Y.}~\bibnamefont {Tang}},
  \bibinfo {author} {\bibfnamefont {J.}~\bibnamefont {Lee}}, \bibinfo {author}
  {\bibfnamefont {J.~H.}\ \bibnamefont {Pixley}}, \bibinfo {author}
  {\bibfnamefont {N.}~\bibnamefont {Andrei}},\ and\ \bibinfo {author}
  {\bibfnamefont {P.}~\bibnamefont {Azaria}},\ }\bibfield  {title} {\bibinfo
  {title} {Spin fractionalization and zero modes in the spin-$\frac{1}{2}$ xxz
  chain with boundary fields},\ }\href {https://doi.org/10.1103/thlq-h58t}
  {\bibfield  {journal} {\bibinfo  {journal} {Phys. Rev. B}\ }\textbf {\bibinfo
  {volume} {112}},\ \bibinfo {pages} {075121} (\bibinfo {year}
  {2025}{\natexlab{a}})}\BibitemShut {NoStop}%
\bibitem [{\citenamefont {Guan}\ \emph {et~al.}(2013)\citenamefont {Guan},
  \citenamefont {Batchelor},\ and\ \citenamefont {Lee}}]{coldatomsBA}%
  \BibitemOpen
  \bibfield  {author} {\bibinfo {author} {\bibfnamefont {X.-W.}\ \bibnamefont
  {Guan}}, \bibinfo {author} {\bibfnamefont {M.~T.}\ \bibnamefont
  {Batchelor}},\ and\ \bibinfo {author} {\bibfnamefont {C.}~\bibnamefont
  {Lee}},\ }\bibfield  {title} {\bibinfo {title} {Fermi gases in one dimension:
  From bethe ansatz to experiments},\ }\href
  {https://doi.org/10.1103/RevModPhys.85.1633} {\bibfield  {journal} {\bibinfo
  {journal} {Rev. Mod. Phys.}\ }\textbf {\bibinfo {volume} {85}},\ \bibinfo
  {pages} {1633} (\bibinfo {year} {2013})}\BibitemShut {NoStop}%
\bibitem [{\citenamefont {Zhang}\ \emph {et~al.}(2018)\citenamefont {Zhang},
  \citenamefont {Zhu}, \citenamefont {Zhao}, \citenamefont {Yan},\ and\
  \citenamefont {Zhu}}]{Zhangcold}%
  \BibitemOpen
  \bibfield  {author} {\bibinfo {author} {\bibfnamefont {D.-W.}\ \bibnamefont
  {Zhang}}, \bibinfo {author} {\bibfnamefont {Y.-Q.}\ \bibnamefont {Zhu}},
  \bibinfo {author} {\bibfnamefont {Y.~X.}\ \bibnamefont {Zhao}}, \bibinfo
  {author} {\bibfnamefont {H.}~\bibnamefont {Yan}},\ and\ \bibinfo {author}
  {\bibfnamefont {S.-L.}\ \bibnamefont {Zhu}},\ }\bibfield  {title} {\bibinfo
  {title} {Topological quantum matter with cold atoms},\ }\href
  {https://doi.org/10.1080/00018732.2019.1594094} {\bibfield  {journal}
  {\bibinfo  {journal} {Advances in Physics}\ }\textbf {\bibinfo {volume}
  {67}},\ \bibinfo {pages} {253} (\bibinfo {year} {2018})},\ \Eprint
  {https://arxiv.org/abs/https://doi.org/10.1080/00018732.2019.1594094}
  {https://doi.org/10.1080/00018732.2019.1594094} \BibitemShut {NoStop}%
\bibitem [{\citenamefont {Potirniche}\ \emph {et~al.}(2017)\citenamefont
  {Potirniche}, \citenamefont {Potter}, \citenamefont {Schleier-Smith},
  \citenamefont {Vishwanath},\ and\ \citenamefont {Yao}}]{SPTcold}%
  \BibitemOpen
  \bibfield  {author} {\bibinfo {author} {\bibfnamefont {I.~D.}\ \bibnamefont
  {Potirniche}}, \bibinfo {author} {\bibfnamefont {A.~C.}\ \bibnamefont
  {Potter}}, \bibinfo {author} {\bibfnamefont {M.}~\bibnamefont
  {Schleier-Smith}}, \bibinfo {author} {\bibfnamefont {A.}~\bibnamefont
  {Vishwanath}},\ and\ \bibinfo {author} {\bibfnamefont {N.~Y.}\ \bibnamefont
  {Yao}},\ }\bibfield  {title} {\bibinfo {title} {Floquet symmetry-protected
  topological phases in cold-atom systems},\ }\href
  {https://doi.org/10.1103/PhysRevLett.119.123601} {\bibfield  {journal}
  {\bibinfo  {journal} {Phys. Rev. Lett.}\ }\textbf {\bibinfo {volume} {119}},\
  \bibinfo {pages} {123601} (\bibinfo {year} {2017})}\BibitemShut {NoStop}%
\bibitem [{\citenamefont {Houck}\ \emph {et~al.}(2012)\citenamefont {Houck},
  \citenamefont {Tureci},\ and\ \citenamefont {Koch}}]{KochpotSC}%
  \BibitemOpen
  \bibfield  {author} {\bibinfo {author} {\bibfnamefont {A.~A.}\ \bibnamefont
  {Houck}}, \bibinfo {author} {\bibfnamefont {H.~E.}\ \bibnamefont {Tureci}},\
  and\ \bibinfo {author} {\bibfnamefont {J.}~\bibnamefont {Koch}},\ }\bibfield
  {title} {\bibinfo {title} {On-chip quantum simulation with superconducting
  circuits},\ }\href {https://doi.org/10.1038/nphys2251} {\bibfield  {journal}
  {\bibinfo  {journal} {Nature Physics}\ }\textbf {\bibinfo {volume} {8}},\
  \bibinfo {pages} {292} (\bibinfo {year} {2012})}\BibitemShut {NoStop}%
\bibitem [{\citenamefont {Smith}\ \emph {et~al.}(2019)\citenamefont {Smith},
  \citenamefont {Kim}, \citenamefont {Pollmann},\ and\ \citenamefont
  {Knolle}}]{Quantumcircuitdynamics}%
  \BibitemOpen
  \bibfield  {author} {\bibinfo {author} {\bibfnamefont {A.}~\bibnamefont
  {Smith}}, \bibinfo {author} {\bibfnamefont {M.~S.}\ \bibnamefont {Kim}},
  \bibinfo {author} {\bibfnamefont {F.}~\bibnamefont {Pollmann}},\ and\
  \bibinfo {author} {\bibfnamefont {J.}~\bibnamefont {Knolle}},\ }\bibfield
  {title} {\bibinfo {title} {Simulating quantum many-body dynamics on a current
  digital quantum computer},\ }\href
  {https://doi.org/10.1038/s41534-019-0217-0} {\bibfield  {journal} {\bibinfo
  {journal} {npj Quantum Information}\ }\textbf {\bibinfo {volume} {5}},\
  \bibinfo {pages} {106} (\bibinfo {year} {2019})}\BibitemShut {NoStop}%
\bibitem [{\citenamefont {Glazman}\ and\ \citenamefont
  {Larkin}(1997)}]{LarkinSC}%
  \BibitemOpen
  \bibfield  {author} {\bibinfo {author} {\bibfnamefont {L.~I.}\ \bibnamefont
  {Glazman}}\ and\ \bibinfo {author} {\bibfnamefont {A.~I.}\ \bibnamefont
  {Larkin}},\ }\bibfield  {title} {\bibinfo {title} {New quantum phase in a
  one-dimensional josephson array},\ }\href
  {https://doi.org/10.1103/PhysRevLett.79.3736} {\bibfield  {journal} {\bibinfo
   {journal} {Phys. Rev. Lett.}\ }\textbf {\bibinfo {volume} {79}},\ \bibinfo
  {pages} {3736} (\bibinfo {year} {1997})}\BibitemShut {NoStop}%
\bibitem [{\citenamefont {Pasnoori}\ \emph
  {et~al.}(2025{\natexlab{b}})\citenamefont {Pasnoori}, \citenamefont
  {Azaria},\ and\ \citenamefont {Mizel}}]{circuitSPT}%
  \BibitemOpen
  \bibfield  {author} {\bibinfo {author} {\bibfnamefont {P.~R.}\ \bibnamefont
  {Pasnoori}}, \bibinfo {author} {\bibfnamefont {P.}~\bibnamefont {Azaria}},\
  and\ \bibinfo {author} {\bibfnamefont {A.}~\bibnamefont {Mizel}},\ }\href
  {https://arxiv.org/abs/2503.13406} {\bibinfo {title} {Realizing a symmetry
  protected topological phase in a superconducting circuit}} (\bibinfo {year}
  {2025}{\natexlab{b}}),\ \Eprint {https://arxiv.org/abs/2503.13406}
  {arXiv:2503.13406 [quant-ph]} \BibitemShut {NoStop}%
\bibitem [{\citenamefont {Bethe}(1931)}]{Bethe1931}%
  \BibitemOpen
  \bibfield  {author} {\bibinfo {author} {\bibfnamefont {H.}~\bibnamefont
  {Bethe}},\ }\bibfield  {title} {\bibinfo {title} {Zur theorie der metalle},\
  }\href {https://doi.org/10.1007/BF01341708} {\bibfield  {journal} {\bibinfo
  {journal} {Zeitschrift fur Physik}\ }\textbf {\bibinfo {volume} {71}},\
  \bibinfo {pages} {205} (\bibinfo {year} {1931})}\BibitemShut {NoStop}%
\bibitem [{\citenamefont {Kondo}(1964)}]{JunKondo}%
  \BibitemOpen
  \bibfield  {author} {\bibinfo {author} {\bibfnamefont {J.}~\bibnamefont
  {Kondo}},\ }\bibfield  {title} {\bibinfo {title} {Resistance minimum in
  dilute magnetic alloys},\ }\href {https://doi.org/10.1143/PTP.32.37}
  {\bibfield  {journal} {\bibinfo  {journal} {Progress of Theoretical Physics}\
  }\textbf {\bibinfo {volume} {32}},\ \bibinfo {pages} {37} (\bibinfo {year}
  {1964})}\BibitemShut {NoStop}%
\bibitem [{\citenamefont {Thirring}(1958)}]{THIRRING}%
  \BibitemOpen
  \bibfield  {author} {\bibinfo {author} {\bibfnamefont {W.~E.}\ \bibnamefont
  {Thirring}},\ }\bibfield  {title} {\bibinfo {title} {A soluble relativistic
  field theory},\ }\href
  {https://doi.org/https://doi.org/10.1016/0003-4916(58)90015-0} {\bibfield
  {journal} {\bibinfo  {journal} {Annals of Physics}\ }\textbf {\bibinfo
  {volume} {3}},\ \bibinfo {pages} {91} (\bibinfo {year} {1958})}\BibitemShut
  {NoStop}%
\bibitem [{\citenamefont {Pasnoori}\ \emph
  {et~al.}(2025{\natexlab{c}})\citenamefont {Pasnoori}, \citenamefont {Mizel},\
  and\ \citenamefont {Azaria}}]{pasnooriduality25}%
  \BibitemOpen
  \bibfield  {author} {\bibinfo {author} {\bibfnamefont {P.~R.}\ \bibnamefont
  {Pasnoori}}, \bibinfo {author} {\bibfnamefont {A.}~\bibnamefont {Mizel}},\
  and\ \bibinfo {author} {\bibfnamefont {P.}~\bibnamefont {Azaria}},\ }\href
  {https://arxiv.org/abs/2503.14776} {\bibinfo {title} {Duality symmetry, zero
  energy modes and boundary spectrum of the sine-gordon/massive thirring
  model}} (\bibinfo {year} {2025}{\natexlab{c}}),\ \Eprint
  {https://arxiv.org/abs/2503.14776} {arXiv:2503.14776 [hep-th]} \BibitemShut
  {NoStop}%
\bibitem [{\citenamefont {Coleman}(1975)}]{ColemanSGMT}%
  \BibitemOpen
  \bibfield  {author} {\bibinfo {author} {\bibfnamefont {S.}~\bibnamefont
  {Coleman}},\ }\bibfield  {title} {\bibinfo {title} {Quantum sine-gordon
  equation as the massive thirring model},\ }\href
  {https://doi.org/10.1103/PhysRevD.11.2088} {\bibfield  {journal} {\bibinfo
  {journal} {Phys. Rev. D}\ }\textbf {\bibinfo {volume} {11}},\ \bibinfo
  {pages} {2088} (\bibinfo {year} {1975})}\BibitemShut {NoStop}%
\bibitem [{\citenamefont {Yang}(1967)}]{Yang}%
  \BibitemOpen
  \bibfield  {author} {\bibinfo {author} {\bibfnamefont {C.~N.}\ \bibnamefont
  {Yang}},\ }\bibfield  {title} {\bibinfo {title} {Some exact results for the
  many-body problem in one dimension with repulsive delta-function
  interaction},\ }\href {https://doi.org/10.1103/PhysRevLett.19.1312}
  {\bibfield  {journal} {\bibinfo  {journal} {Phys. Rev. Lett.}\ }\textbf
  {\bibinfo {volume} {19}},\ \bibinfo {pages} {1312} (\bibinfo {year}
  {1967})}\BibitemShut {NoStop}%
\bibitem [{\citenamefont {Baxter}(1972)}]{BAXTER}%
  \BibitemOpen
  \bibfield  {author} {\bibinfo {author} {\bibfnamefont {R.~J.}\ \bibnamefont
  {Baxter}},\ }\bibfield  {title} {\bibinfo {title} {Partition function of the
  eight-vertex lattice model},\ }\href
  {https://doi.org/https://doi.org/10.1016/0003-4916(72)90335-1} {\bibfield
  {journal} {\bibinfo  {journal} {Annals of Physics}\ }\textbf {\bibinfo
  {volume} {70}},\ \bibinfo {pages} {193} (\bibinfo {year} {1972})}\BibitemShut
  {NoStop}%
\bibitem [{\citenamefont {Zamolodchikov}\ and\ \citenamefont
  {Zamolodchikov}(1979)}]{ZAMOLODCHIKOV}%
  \BibitemOpen
  \bibfield  {author} {\bibinfo {author} {\bibfnamefont {A.~B.}\ \bibnamefont
  {Zamolodchikov}}\ and\ \bibinfo {author} {\bibfnamefont {A.~B.}\ \bibnamefont
  {Zamolodchikov}},\ }\bibfield  {title} {\bibinfo {title} {Factorized
  s-matrices in two dimensions as the exact solutions of certain relativistic
  quantum field theory models},\ }\href
  {https://doi.org/https://doi.org/10.1016/0003-4916(79)90391-9} {\bibfield
  {journal} {\bibinfo  {journal} {Annals of Physics}\ }\textbf {\bibinfo
  {volume} {120}},\ \bibinfo {pages} {253} (\bibinfo {year}
  {1979})}\BibitemShut {NoStop}%
\bibitem [{\citenamefont {Sklyanin}\ \emph {et~al.}(1979)\citenamefont
  {Sklyanin}, \citenamefont {Takhtadzhyan},\ and\ \citenamefont
  {Faddeev}}]{SklyaninQISM}%
  \BibitemOpen
  \bibfield  {author} {\bibinfo {author} {\bibfnamefont {E.~K.}\ \bibnamefont
  {Sklyanin}}, \bibinfo {author} {\bibfnamefont {L.~A.}\ \bibnamefont
  {Takhtadzhyan}},\ and\ \bibinfo {author} {\bibfnamefont {L.~D.}\ \bibnamefont
  {Faddeev}},\ }\bibfield  {title} {\bibinfo {title} {Quantum inverse problem
  method 1},\ }\href
  {http://inis.iaea.org/search/search.aspx?orig_q=RN:11506395} {\bibfield
  {journal} {\bibinfo  {journal} {Teoreticheskaya i Matematicheskaya Fizika}\
  }\bibinfo {series} {Kvantovyj metod obratnoj zadachi 1},\ \textbf {\bibinfo
  {volume} {40}},\ \bibinfo {pages} {194} (\bibinfo {year} {1979})}\BibitemShut
  {NoStop}%
\bibitem [{\citenamefont {Japaridze}\ \emph {et~al.}(1984)\citenamefont
  {Japaridze}, \citenamefont {Nersesyan},\ and\ \citenamefont
  {Wiegmann}}]{Japaridze}%
  \BibitemOpen
  \bibfield  {author} {\bibinfo {author} {\bibfnamefont {G.}~\bibnamefont
  {Japaridze}}, \bibinfo {author} {\bibfnamefont {A.}~\bibnamefont
  {Nersesyan}},\ and\ \bibinfo {author} {\bibfnamefont {P.}~\bibnamefont
  {Wiegmann}},\ }\bibfield  {title} {\bibinfo {title} {Exact results in the
  two-dimensional u(1)-symmetric thirring model},\ }\href@noop {} {\bibfield
  {journal} {\bibinfo  {journal} {Nucl. Phys. B}\ }\textbf {\bibinfo {volume}
  {230}},\ \bibinfo {pages} {511} (\bibinfo {year} {1984})}\BibitemShut
  {NoStop}%
\bibitem [{\citenamefont {Bergknoff}\ and\ \citenamefont
  {Thacker}(1979)}]{Thacker}%
  \BibitemOpen
  \bibfield  {author} {\bibinfo {author} {\bibfnamefont {H.}~\bibnamefont
  {Bergknoff}}\ and\ \bibinfo {author} {\bibfnamefont {H.~B.}\ \bibnamefont
  {Thacker}},\ }\bibfield  {title} {\bibinfo {title} {Structure and solution of
  the massive thirring model},\ }\href
  {https://doi.org/10.1103/PhysRevD.19.3666} {\bibfield  {journal} {\bibinfo
  {journal} {Phys. Rev. D}\ }\textbf {\bibinfo {volume} {19}},\ \bibinfo
  {pages} {3666} (\bibinfo {year} {1979})}\BibitemShut {NoStop}%
\bibitem [{\citenamefont {Richardson}\ and\ \citenamefont
  {Sherman}(1964)}]{RICHARDSON}%
  \BibitemOpen
  \bibfield  {author} {\bibinfo {author} {\bibfnamefont {R.}~\bibnamefont
  {Richardson}}\ and\ \bibinfo {author} {\bibfnamefont {N.}~\bibnamefont
  {Sherman}},\ }\bibfield  {title} {\bibinfo {title} {Exact eigenstates of the
  pairing-force hamiltonian},\ }\href
  {https://doi.org/https://doi.org/10.1016/0029-5582(64)90687-X} {\bibfield
  {journal} {\bibinfo  {journal} {Nuclear Physics}\ }\textbf {\bibinfo {volume}
  {52}},\ \bibinfo {pages} {221} (\bibinfo {year} {1964})}\BibitemShut
  {NoStop}%
\bibitem [{\citenamefont {Dukelsky}\ \emph {et~al.}(2004)\citenamefont
  {Dukelsky}, \citenamefont {Pittel},\ and\ \citenamefont {Sierra}}]{Dukelsky}%
  \BibitemOpen
  \bibfield  {author} {\bibinfo {author} {\bibfnamefont {J.}~\bibnamefont
  {Dukelsky}}, \bibinfo {author} {\bibfnamefont {S.}~\bibnamefont {Pittel}},\
  and\ \bibinfo {author} {\bibfnamefont {G.}~\bibnamefont {Sierra}},\
  }\bibfield  {title} {\bibinfo {title} {Colloquium: Exactly solvable
  richardson-gaudin models for many-body quantum systems},\ }\href
  {https://doi.org/10.1103/RevModPhys.76.643} {\bibfield  {journal} {\bibinfo
  {journal} {Rev. Mod. Phys.}\ }\textbf {\bibinfo {volume} {76}},\ \bibinfo
  {pages} {643} (\bibinfo {year} {2004})}\BibitemShut {NoStop}%
\bibitem [{\citenamefont {{Gaudin, M.}}(1976)}]{Gaudin1976}%
  \BibitemOpen
  \bibfield  {author} {\bibinfo {author} {\bibnamefont {{Gaudin, M.}}},\
  }\bibfield  {title} {\bibinfo {title} {Diagonalisation d'une classe
  d'hamiltoniens de spin},\ }\href
  {https://doi.org/10.1051/jphys:0197600370100108700} {\bibfield  {journal}
  {\bibinfo  {journal} {J. Phys. France}\ }\textbf {\bibinfo {volume} {37}},\
  \bibinfo {pages} {1087} (\bibinfo {year} {1976})}\BibitemShut {NoStop}%
\bibitem [{\citenamefont {Raizen}(1999)}]{RAIZENcold}%
  \BibitemOpen
  \bibfield  {author} {\bibinfo {author} {\bibfnamefont {M.~G.}\ \bibnamefont
  {Raizen}},\ }\bibfield  {title} {\bibinfo {title} {Quantum chaos with cold
  atoms}\ }(\bibinfo  {publisher} {Academic Press},\ \bibinfo {year} {1999})\
  pp.\ \bibinfo {pages} {43--81}\BibitemShut {NoStop}%
\bibitem [{\citenamefont {Siva}\ \emph {et~al.}(2023)\citenamefont {Siva},
  \citenamefont {Koolstra}, \citenamefont {Steinmetz}, \citenamefont
  {Livingston}, \citenamefont {Das}, \citenamefont {Chen}, \citenamefont
  {Kreikebaum}, \citenamefont {Stevenson}, \citenamefont {Junger},
  \citenamefont {Santiago}, \citenamefont {Siddiqi},\ and\ \citenamefont
  {Jordan}}]{circuitHamiltonian}%
  \BibitemOpen
  \bibfield  {author} {\bibinfo {author} {\bibfnamefont {K.}~\bibnamefont
  {Siva}}, \bibinfo {author} {\bibfnamefont {G.}~\bibnamefont {Koolstra}},
  \bibinfo {author} {\bibfnamefont {J.}~\bibnamefont {Steinmetz}}, \bibinfo
  {author} {\bibfnamefont {W.~P.}\ \bibnamefont {Livingston}}, \bibinfo
  {author} {\bibfnamefont {D.}~\bibnamefont {Das}}, \bibinfo {author}
  {\bibfnamefont {L.}~\bibnamefont {Chen}}, \bibinfo {author} {\bibfnamefont
  {J.}~\bibnamefont {Kreikebaum}}, \bibinfo {author} {\bibfnamefont
  {N.}~\bibnamefont {Stevenson}}, \bibinfo {author} {\bibfnamefont
  {C.}~\bibnamefont {Junger}}, \bibinfo {author} {\bibfnamefont
  {D.}~\bibnamefont {Santiago}}, \bibinfo {author} {\bibfnamefont
  {I.}~\bibnamefont {Siddiqi}},\ and\ \bibinfo {author} {\bibfnamefont
  {A.}~\bibnamefont {Jordan}},\ }\bibfield  {title} {\bibinfo {title}
  {Time-dependent hamiltonian reconstruction using continuous weak
  measurements},\ }\href {https://doi.org/10.1103/PRXQuantum.4.040324}
  {\bibfield  {journal} {\bibinfo  {journal} {PRX Quantum}\ }\textbf {\bibinfo
  {volume} {4}},\ \bibinfo {pages} {040324} (\bibinfo {year}
  {2023})}\BibitemShut {NoStop}%
\bibitem [{\citenamefont {Yuzbashyan}(2018)}]{YUZBASHYAN}%
  \BibitemOpen
  \bibfield  {author} {\bibinfo {author} {\bibfnamefont {E.~A.}\ \bibnamefont
  {Yuzbashyan}},\ }\bibfield  {title} {\bibinfo {title} {Integrable
  time-dependent hamiltonians, solvable landau-zener models and gaudin
  magnets},\ }\href {https://doi.org/https://doi.org/10.1016/j.aop.2018.01.017}
  {\bibfield  {journal} {\bibinfo  {journal} {Annals of Physics}\ }\textbf
  {\bibinfo {volume} {392}},\ \bibinfo {pages} {323} (\bibinfo {year}
  {2018})}\BibitemShut {NoStop}%
\bibitem [{\citenamefont {Demkov}\ and\ \citenamefont
  {Osherov}(1968)}]{demkov1968stationary}%
  \BibitemOpen
  \bibfield  {author} {\bibinfo {author} {\bibfnamefont {Y.~N.}\ \bibnamefont
  {Demkov}}\ and\ \bibinfo {author} {\bibfnamefont {V.}~\bibnamefont
  {Osherov}},\ }\bibfield  {title} {\bibinfo {title} {Stationary and
  nonstationary problems in quantum mechanics that can be solved by means of
  contour integration},\ }\href@noop {} {\bibfield  {journal} {\bibinfo
  {journal} {Sov. Phys. JETP}\ }\textbf {\bibinfo {volume} {26}},\ \bibinfo
  {pages} {1} (\bibinfo {year} {1968})}\BibitemShut {NoStop}%
\bibitem [{\citenamefont {Patra}\ and\ \citenamefont
  {Yuzbashyan}(2015)}]{Patra_2015}%
  \BibitemOpen
  \bibfield  {author} {\bibinfo {author} {\bibfnamefont {A.}~\bibnamefont
  {Patra}}\ and\ \bibinfo {author} {\bibfnamefont {E.~A.}\ \bibnamefont
  {Yuzbashyan}},\ }\bibfield  {title} {\bibinfo {title} {Quantum integrability
  in the multistate landau-zener problem},\ }\href
  {https://doi.org/10.1088/1751-8113/48/24/245303} {\bibfield  {journal}
  {\bibinfo  {journal} {Journal of Physics A: Mathematical and Theoretical}\
  }\textbf {\bibinfo {volume} {48}},\ \bibinfo {pages} {245303} (\bibinfo
  {year} {2015})}\BibitemShut {NoStop}%
\bibitem [{\citenamefont {Anderson}\ \emph {et~al.}(1970)\citenamefont
  {Anderson}, \citenamefont {Yuval},\ and\ \citenamefont {Hamann}}]{Anderson}%
  \BibitemOpen
  \bibfield  {author} {\bibinfo {author} {\bibfnamefont {P.~W.}\ \bibnamefont
  {Anderson}}, \bibinfo {author} {\bibfnamefont {G.}~\bibnamefont {Yuval}},\
  and\ \bibinfo {author} {\bibfnamefont {D.~R.}\ \bibnamefont {Hamann}},\
  }\bibfield  {title} {\bibinfo {title} {Exact results in the kondo problem.
  ii. scaling theory, qualitatively correct solution, and some new results on
  one-dimensional classical statistical models},\ }\href
  {https://doi.org/10.1103/PhysRevB.1.4464} {\bibfield  {journal} {\bibinfo
  {journal} {Phys. Rev. B}\ }\textbf {\bibinfo {volume} {1}},\ \bibinfo {pages}
  {4464} (\bibinfo {year} {1970})}\BibitemShut {NoStop}%
\bibitem [{\citenamefont {Wilson}(1975)}]{Wilson}%
  \BibitemOpen
  \bibfield  {author} {\bibinfo {author} {\bibfnamefont {K.~G.}\ \bibnamefont
  {Wilson}},\ }\bibfield  {title} {\bibinfo {title} {The renormalization group:
  Critical phenomena and the kondo problem},\ }\href
  {https://doi.org/10.1103/RevModPhys.47.773} {\bibfield  {journal} {\bibinfo
  {journal} {Rev. Mod. Phys.}\ }\textbf {\bibinfo {volume} {47}},\ \bibinfo
  {pages} {773} (\bibinfo {year} {1975})}\BibitemShut {NoStop}%
\bibitem [{\citenamefont {Nozieres}\ and\ \citenamefont
  {Blandin}(1980)}]{Nozier}%
  \BibitemOpen
  \bibfield  {author} {\bibinfo {author} {\bibfnamefont {P.}~\bibnamefont
  {Nozieres}}\ and\ \bibinfo {author} {\bibfnamefont {A.}~\bibnamefont
  {Blandin}},\ }\bibfield  {title} {\bibinfo {title} {Kondo effect in real
  metals},\ }\href {https://doi.org/10.1051/jphys:01980004103019300} {\bibfield
   {journal} {\bibinfo  {journal} {J. Phys. France}\ }\textbf {\bibinfo
  {volume} {41}},\ \bibinfo {pages} {193} (\bibinfo {year} {1980})}\BibitemShut
  {NoStop}%
\bibitem [{\citenamefont {Andrei}(1980)}]{Andrei80}%
  \BibitemOpen
  \bibfield  {author} {\bibinfo {author} {\bibfnamefont {N.}~\bibnamefont
  {Andrei}},\ }\bibfield  {title} {\bibinfo {title} {Diagonalization of the
  kondo hamiltonian},\ }\href {https://doi.org/10.1103/PhysRevLett.45.379}
  {\bibfield  {journal} {\bibinfo  {journal} {Phys. Rev. Lett.}\ }\textbf
  {\bibinfo {volume} {45}},\ \bibinfo {pages} {379} (\bibinfo {year}
  {1980})}\BibitemShut {NoStop}%
\bibitem [{\citenamefont {Wiegmann}(1981)}]{Wiegmann_1981}%
  \BibitemOpen
  \bibfield  {author} {\bibinfo {author} {\bibfnamefont {P.~B.}\ \bibnamefont
  {Wiegmann}},\ }\bibfield  {title} {\bibinfo {title} {Exact solution of the
  s-d exchange model (kondo problem)},\ }\href
  {https://doi.org/10.1088/0022-3719/14/10/014} {\bibfield  {journal} {\bibinfo
   {journal} {Journal of Physics C: Solid State Physics}\ }\textbf {\bibinfo
  {volume} {14}},\ \bibinfo {pages} {1463} (\bibinfo {year}
  {1981})}\BibitemShut {NoStop}%
\bibitem [{\citenamefont {Pasnoori}(2025)}]{Supplemental}%
  \BibitemOpen
  \bibfield  {author} {\bibinfo {author} {\bibfnamefont {P.~R.}\ \bibnamefont
  {Pasnoori}},\ }\href@noop {} {\bibinfo {title} {Supplemental material [link
  to be inserted by publisher]}} (\bibinfo {year} {2025})\BibitemShut {NoStop}%
\bibitem [{Note1()}]{Note1}%
  \BibitemOpen
  \bibinfo {note} {Explicit form of the permutation operator is \begin
  {equation}P^{ab}_{pq,rs}=\protect \frac {1}{2}\left (I^{ab}_{pq,rs}+\DOTSB
  \sum@ \slimits@ _{\alpha =x,y,z}\sigma ^{a,\alpha }_{pq}\sigma ^{b,\alpha
  }_{rs}\right ). \end {equation}Here $\sigma ^{a,\alpha }$ and $\sigma
  ^{b,\alpha }$, $\alpha =x,y,z$ are the Pauli matrices acting in the spin
  spaces $a$ and $b$ respectively.}\BibitemShut {Stop}%
\bibitem [{Note2()}]{Note2}%
  \BibitemOpen
  \bibinfo {note} {In the case of quadratic dispersion, the Hamiltonian fixes
  all the relations and this is one of the reasons why the regular Kondo model
  is not integrable with quadratic dispersion.}\BibitemShut {Stop}%
\bibitem [{Note3()}]{Note3}%
  \BibitemOpen
  \bibinfo {note} {These conditions impose that transporting particle `j'
  around the system first and then particle `i' around the system is equivalent
  to transporting particle `i' around the system first and then particle `j'
  around the system.}\BibitemShut {Stop}%
\bibitem [{\citenamefont {Ruijsenaars}(1997)}]{ruijisenars}%
  \BibitemOpen
  \bibfield  {author} {\bibinfo {author} {\bibfnamefont {S.~N.~M.}\
  \bibnamefont {Ruijsenaars}},\ }\bibfield  {title} {\bibinfo {title} {First
  order analytic difference equations and integrable quantum systems},\ }\href
  {https://doi.org/10.1063/1.531809} {\bibfield  {journal} {\bibinfo  {journal}
  {Journal of Mathematical Physics}\ }\textbf {\bibinfo {volume} {38}},\
  \bibinfo {pages} {1069} (\bibinfo {year} {1997})},\ \Eprint
  {https://arxiv.org/abs/https://pubs.aip.org/aip/jmp/article-pdf/38/2/1069/19136488/1069\_1\_online.pdf}
  {https://pubs.aip.org/aip/jmp/article-pdf/38/2/1069/19136488/1069\_1\_online.pdf}
  \BibitemShut {NoStop}%
\bibitem [{Note4()}]{Note4}%
  \BibitemOpen
  \bibinfo {note} {The case where $a,c \in \protect \mathbb {C}$ is also a
  solution, in which case, the Hamiltonian is non Hermitian and is interesting
  on its own right.}\BibitemShut {Stop}%
\bibitem [{\citenamefont {Reshetikhin}(1992{\natexlab{a}})}]{rishetikhin1}%
  \BibitemOpen
  \bibfield  {author} {\bibinfo {author} {\bibfnamefont {N.}~\bibnamefont
  {Reshetikhin}},\ }\bibfield  {title} {\bibinfo {title} {Jackson-type
  integrals, bethe vectors, and solutions to a difference analog of the
  knizhnik-zamolodchikov system},\ }\href {https://doi.org/10.1007/BF00420749}
  {\bibfield  {journal} {\bibinfo  {journal} {Letters in Mathematical Physics}\
  }\textbf {\bibinfo {volume} {26}},\ \bibinfo {pages} {153} (\bibinfo {year}
  {1992}{\natexlab{a}})}\BibitemShut {NoStop}%
\bibitem [{\citenamefont {Smirnov}(1986)}]{Smirnov_1986}%
  \BibitemOpen
  \bibfield  {author} {\bibinfo {author} {\bibfnamefont {F.~A.}\ \bibnamefont
  {Smirnov}},\ }\bibfield  {title} {\bibinfo {title} {A general formula for
  soliton form factors in the quantum sine-gordon model},\ }\href
  {https://doi.org/10.1088/0305-4470/19/10/003} {\bibfield  {journal} {\bibinfo
   {journal} {Journal of Physics A: Mathematical and General}\ }\textbf
  {\bibinfo {volume} {19}},\ \bibinfo {pages} {L575} (\bibinfo {year}
  {1986})}\BibitemShut {NoStop}%
\bibitem [{\citenamefont {Frenkel}\ and\ \citenamefont
  {Reshetikhin}(1992)}]{Frenkel}%
  \BibitemOpen
  \bibfield  {author} {\bibinfo {author} {\bibfnamefont {I.~B.}\ \bibnamefont
  {Frenkel}}\ and\ \bibinfo {author} {\bibfnamefont {N.~Y.}\ \bibnamefont
  {Reshetikhin}},\ }\bibfield  {title} {\bibinfo {title} {Quantum affine
  algebras and holonomic difference equations},\ }\href
  {https://doi.org/10.1007/BF02099206} {\bibfield  {journal} {\bibinfo
  {journal} {Communications in Mathematical Physics}\ }\textbf {\bibinfo
  {volume} {146}},\ \bibinfo {pages} {1} (\bibinfo {year} {1992})}\BibitemShut
  {NoStop}%
\bibitem [{\citenamefont {Reshetikhin}(1992{\natexlab{b}})}]{rishetikhin2}%
  \BibitemOpen
  \bibfield  {author} {\bibinfo {author} {\bibfnamefont {N.}~\bibnamefont
  {Reshetikhin}},\ }\bibfield  {title} {\bibinfo {title} {The
  knizhnik-zamolodchikov system as a deformation of the isomonodromy problem},\
  }\href {https://doi.org/10.1007/BF00420750} {\bibfield  {journal} {\bibinfo
  {journal} {Letters in Mathematical Physics}\ }\textbf {\bibinfo {volume}
  {26}},\ \bibinfo {pages} {167} (\bibinfo {year}
  {1992}{\natexlab{b}})}\BibitemShut {NoStop}%
\bibitem [{\citenamefont {Varchenko}(1995)}]{varchenko}%
  \BibitemOpen
  \bibfield  {author} {\bibinfo {author} {\bibfnamefont {A.~N.}\ \bibnamefont
  {Varchenko}},\ }\bibfield  {title} {\bibinfo {title} {Asymptotic solutions to
  the knizhnik-zamolodchikov equation and crystal base},\ }\href
  {https://doi.org/10.1007/BF02103772} {\bibfield  {journal} {\bibinfo
  {journal} {Communications in Mathematical Physics}\ }\textbf {\bibinfo
  {volume} {171}},\ \bibinfo {pages} {99} (\bibinfo {year} {1995})}\BibitemShut
  {NoStop}%
\bibitem [{\citenamefont {Matsuo}(1993)}]{matsuo}%
  \BibitemOpen
  \bibfield  {author} {\bibinfo {author} {\bibfnamefont {A.}~\bibnamefont
  {Matsuo}},\ }\bibfield  {title} {\bibinfo {title} {Jackson integrals of
  jordan-pochhammer type and quantum knizhnik-zamolodchikov equations},\ }\href
  {https://doi.org/10.1007/BF02096769} {\bibfield  {journal} {\bibinfo
  {journal} {Communications in Mathematical Physics}\ }\textbf {\bibinfo
  {volume} {151}},\ \bibinfo {pages} {263} (\bibinfo {year}
  {1993})}\BibitemShut {NoStop}%
\bibitem [{\citenamefont {Babujian}\ \emph {et~al.}(1997)\citenamefont
  {Babujian}, \citenamefont {Karowski},\ and\ \citenamefont
  {Zapletal}}]{Babujian_1997}%
  \BibitemOpen
  \bibfield  {author} {\bibinfo {author} {\bibfnamefont {H.}~\bibnamefont
  {Babujian}}, \bibinfo {author} {\bibfnamefont {M.}~\bibnamefont {Karowski}},\
  and\ \bibinfo {author} {\bibfnamefont {A.}~\bibnamefont {Zapletal}},\
  }\bibfield  {title} {\bibinfo {title} {Matrix difference equations and a
  nested bethe ansatz},\ }\href {https://doi.org/10.1088/0305-4470/30/18/019}
  {\bibfield  {journal} {\bibinfo  {journal} {Journal of Physics A:
  Mathematical and General}\ }\textbf {\bibinfo {volume} {30}},\ \bibinfo
  {pages} {6425} (\bibinfo {year} {1997})}\BibitemShut {NoStop}%
\bibitem [{\citenamefont {Pasnoori}\ and\ \citenamefont
  {Yuzbashyan}()}]{ParmeshEmil}%
  \BibitemOpen
  \bibfield  {author} {\bibinfo {author} {\bibfnamefont {P.~R.~.}\ \bibnamefont
  {Pasnoori}}\ and\ \bibinfo {author} {\bibfnamefont {E.~A.}\ \bibnamefont
  {Yuzbashyan}},\ }\href@noop {} {\bibinfo  {journal} {In preparation}\
  }\BibitemShut {NoStop}%
\bibitem [{\citenamefont {Fendley}(2016)}]{Fendley}%
  \BibitemOpen
\bibfield  {journal} {  }\bibfield  {author} {\bibinfo {author} {\bibfnamefont
  {P.}~\bibnamefont {Fendley}},\ }\bibfield  {title} {\bibinfo {title} {Strong
  zero modes and eigenstate phase transitions in the {XYZ}/interacting majorana
  chain},\ }\href {https://doi.org/10.1088/1751-8113/49/30/30lt01} {\bibfield
  {journal} {\bibinfo  {journal} {Journal of Physics A: Mathematical and
  Theoretical}\ }\textbf {\bibinfo {volume} {49}},\ \bibinfo {pages} {30LT01}
  (\bibinfo {year} {2016})}\BibitemShut {NoStop}%
\end{thebibliography}%

\begin{widetext}

\section{One particle wavefunction}
In this section we construct the most general one particle wavefunction presented in the main text. The Hamiltonian is given by

\bea \nonumber H=\int_{0}^{L} \; dx\;  \big\{\Psi^{\dagger}_{a}(x)(-i\partial_x)\Psi_{a}(x) + J(t) \Psi^{\dagger}_{a}(0)\left(\vec{\sigma}_{ab}\cdot\vec{S}_{\alpha\beta}\right)\Psi_{b}(0)\big\}, \label{Hamiltonianap}
\eea

where $\Psi_{a}(x)$ describes the fermion (electron) field with subscript $a=\uparrow,\downarrow$ denoting the spin. $S$ represents the impurity and $J(t)$ is the time dependent interaction strength. Since the system conserves the total number of electrons $N$
\bea N=\sum_{a}\int_{0}^{L} dx \Psi^{\dagger}_{a}(x)\Psi_{a}(x), 
\eea

we can look for wave function labeled by $N$ which satisfies the time dependent Schrodinger equation

\bea i\partial_t \ket{\Psi_N}= H \ket{\Psi_N}.\label{SEkap}\eea

The wave function in the one particle sector can be written as
\bea \label{1pformkap}\ket{\Psi_1}=\sum_{a\alpha} \int_0^L dx \; \Psi^{\dagger}_a(x) F_{a\alpha}(x,t)\ket{\alpha}.
\eea

Here $a,\alpha$ denote the spin degrees of freedom of the particle and the impurity respectively. Using the above expression in the Schrodinger equation (\ref{SEkap}), we obtain 
\begin{align}-i(\partial_t+\partial_x)F_{a\alpha}(x,t)+J(t) \delta(x) \left(\vec{\sigma}_{ab}\cdot\vec{S}_{\alpha\beta}\right)F_{b\beta}(x,t)=0.\end{align}

\subsubsection{A simple wavefunction for time dependent interaction strength $J(t)$}
To gain some intuition, let us consider the situation where the system is very large such that we can ignore the boundary conditions. Let the wave function associated with the particle be sharply localized in the form of a wave packet 
\be \label{1p1ap}F_{a\alpha}(x,t)= f_{a\alpha}(-\tau)\:e^{-\frac{1}{2\sigma^2}\left(x-t+\tau\right)^2} \;\; (\text{in the absence of impurity}) \ee 

with very small variance $\sigma\ll 1$. Here $f_{a\alpha}(-\tau)$ is an amplitude with $a$ and $\alpha$ being the spin indices of the particle and the impurity respectively. The reason for the negative sign in the argument of $f_{a\alpha}(-\tau)$ is for notational ease, as will be evident below. The wave packet moves to the right at the speed of $v_F$, which we have set to 1. In the absence of the impurity, the Schrodinger equation (\ref{SEkap}) only has the first term and the wavefunction (\ref{1p1ap}) is a solution. Note that since we have linear dispersion, the shape of the wavefunction does not change, but only propagates to the right. 

\vspace{3mm}

Now let us consider the situation in which the impurity is present. It is clear that the wave packet is on the left side of the impurity for $t<\tau$ and is on the right side of the impurity for $t>\tau$. Since the particle interacts with the impurity through spin exchange interaction, the amplitude $f_{a\alpha}(-\tau)$ will be different on either side of the impurity. Hence, the wavefunction now takes the form

\be \label{1p2ap}F_{a\alpha}(x,t)=\left(f^{10}_{a\alpha}(-\tau)\theta(-x)+f^{01}_{a\alpha}(-\tau)\theta(x)\right)e^{-\frac{1}{2\sigma^2}\left(x-t+\tau\right)^2}.\ee

Here $f^{10}_{a\alpha}(-\tau),f^{01}_{a\alpha}(-\tau)$ correspond to the amplitudes when the particle is on the left and right sides of the impurity respectively. $\theta(x)$ is the Heaviside function with the convention $\theta(x)=1, x>0$, $\theta(x)=0, x<0$ and $\theta(0)=1/2$. Using (\ref{1p2ap}) in the Schrodinger equation (\ref{SEkap}) we find that it is a solution when 

\bea f^{01}_{a,\alpha}(-\tau)=S_{ab,\alpha\beta}(-\tau)f^{10}_{b,\beta}(-\tau),
\eea

where the S-matrix $S_{ab,\alpha\beta}(-\tau)$ is given by

\begin{align}
\nonumber & S_{ab,\alpha\beta}(-\tau)= \frac{i g(-\tau) I_{ab,\alpha\beta}+P_{ab,\alpha\beta}}{i g(-\tau)+1} e^{i\phi(-\tau)},\\\nonumber & g(-\tau)= \frac{1}{2 J(\tau)}\left(1-\frac{3 J(\tau)^2}{4}\right),\\ & e^{i\phi(-\tau)}=\frac{1+\left(i/2J(\tau)\right)\left(1-(3/4)J^2(\tau)\right)}{iJ(\tau)-\left(1+(3/4)J^2(\tau)\right)}.\label{smatimp1ap}
\end{align}

where $I, P$ are identity and permutation operators respectively. Hence, we find that the S-matrix between the particle and the impurity which relates the amplitudes associated with the particle on either sides of the impurity is a function of $J(\tau)$, which is the strength of the impurity interaction strength when the wave packet is localized at the impurity.

\subsubsection{General wavefunction for constant $J(t)=J$}
We have considered the situation where the particle is sharply localized in the form of a wave packet. Even though this produces a consistent solution, it clearly cannot be a general solution. To obtain the most general solution, it is natural to consider the superposition of the wave packets with all values of the parameter $\tau$. In other words, at any given time, the wave function we are looking for should be spread out in space instead of being localized like a wave packet considered above. In the case where the impurity interaction strength is constant, one can choose $f^{10}(-\tau)=e^{-ik\tau}f^{10}$ and $f^{01}(-\tau)=e^{-ik\tau}f^{01}$ and consider a superposition of the wave packets for all values of $\tau$ as follows

\be \label{1p3ap}F_{a\alpha}(x,t)=\int_{\tau}\left(f^{10}_{a\alpha}e^{-ik\tau}\theta(-x)+f^{01}_{a\alpha}e^{-ik\tau}\theta(x)\right)e^{-\frac{1}{2\sigma^2}\left(x-t+\tau\right)^2}=\left(f^{10}_{a\alpha}\theta(-x)+f^{01}_{a\alpha}\theta(x)\right)e^{ik(x-t)}.\ee

Using this in the Schrodinger equation, one obtains 

\be f^{01}_{a\alpha}=S_{ab,\alpha\beta} f^{10}_{b\beta}, \;\;S_{ab,\alpha\beta}= \frac{i g \:I_{ab,\alpha\beta}+P_{ab,\alpha\beta}}{i g+1} e^{i\phi(\tau)}, \;\; (\text{constant $J(t)=J$})\ee
where
\be g= \frac{1}{2 J}\left(1-\frac{3 J^2}{4}\right),\;\; e^{i\phi}=\frac{1+\left(i/2J\right)\left(1-(3/4)J^2\right)}{iJ-\left(1+(3/4)J^2\right)}.\ee

This is the usual solution in terms of the plane waves for the model with constant interaction strength.

\subsubsection{General wavefunction for time dependent interaction strength $J(t)$}

Now for general time dependent case $J(t)$, one can consider the superposition of the wave packets for all values of $\tau$ in (\ref{1p2ap})

\begin{align} \label{1p2spap}F_{a\alpha}(x,t)&=\int_{\tau}\left(f^{10}_{a\alpha}(-\tau)\theta(-x)+f^{01}_{a\alpha}(-\tau)\theta(x)\right)e^{-\frac{1}{2\sigma^2}\left(x-t+\tau\right)^2}\\&=f^{10}_{a\alpha}(x-t)\theta(-x)+f^{01}_{a\alpha}(x-t)\theta(x).\label{1p2spap2e}\end{align}

Hence, in the case where the interaction strength is time dependent, we see that the amplitudes corresponding to the particle being on the left and right sides of the impurity are dependent on both the $x$ and time $t$, as opposed to being constant, as in the case where the interaction strength is time independent. The relation between the amplitudes on either sides of the impurity now takes the form

\bea f^{01}_{a,\alpha}(x-t)=S_{ab,\alpha\beta}(x-t)f^{10}_{b,\beta}(x-t),\label{1psmatrelap}
\eea
where $S_{ab,\alpha\beta}(x-t)$ is given by
\begin{align} \label{smatk1ap}&S^{10}_{ab,\alpha\beta}(x-t)=e^{i\phi(x-t)}\frac{ig(z)I^{10}_{ab,\alpha\beta}+P^{10}_{ab,\alpha\beta}}{ig(z)+1},\\
&g(x-t)= \frac{1}{2J(t-x)}\left(1-\frac{3}{4}(J(t-x))^2\right),\\&e^{i\phi(x-t)}=\frac{1+\left(i/2J(t-x)\right)\left(1-(3/4)J^2(t-x)\right)}{iJ(t-x)-\left(1+(3/4)J^2(t-x)\right)}.
\end{align}
In expression (\ref{1p2spap2e}), the amplitude $f^{10}(x-t)$ is `physical' when $x<0$, and `unphysical' for $x>0$. similarly, the amplitude $f^{01}(x-t)$ is physical when $x>0$ and unphysical for $x<0$. The relation (\ref{1psmatrelap}) means that for $x>0$, the amplitude $f^{01}(x-t)$, which is physical, is related to the unphysical amplitude $f^{10}(x-t)$ when multiplied by the operator $S(x-t)$. Equivalently, for for $x<0$, the amplitude $f^{10}(x-t)$, which is physical, is related to the unphysical amplitude $f^{01}(x-t)$ when multiplied by the operator $\left(S(x-t)\right)^{-1}$. Here $\left(S(x-t)\right)^{-1}$ is the inverse of $S^{10}(x-t)$.
\vspace{3mm}

Up until now, we have deliberately ignored the boundary conditions by considering a system of infinite size for the sake of simplicity. To obtain the exact wavefunction, one needs to impose proper boundary conditions on the fermion fields $\Psi_a(x)$ by considering the system with finite size $L$. For the sake of generality, we consider the system to have length $y$ for $x>0$ and $L-y$ for $x<0$. Here $0<y<L$ is an arbitrary constant, which is usually taken to be $y=L/2$, such that the impurity is at the center of the system. Here we choose $y$ to be an arbitrary constant which does not effect the solution in anyway. The most general wavefunction (\ref{1p2spap}) now takes the form provided in the maintext, which is

\begin{align} F_{a\alpha}(x,t)=  \big(f^{10}_{a\alpha}(x-t)\theta(-x)\theta(L-y+x) +f^{01}_{a\alpha}(x-t)\theta(x)\theta(y-x)\big) \;\; \text{ (most general one particle wavefunction)}.\label{1pformkap}
\end{align}

Using this in the Scrodinger equation (\ref{SEkap}) and applying periodic boundary conditions on the fermion fields $\Psi_a(y)=\Psi_a(L-y)$, we obtain the following relation provided in the main text

\be f^{10}_{a\alpha}(x-t-L)=f^{01}_{a\alpha}(x-t).\label{1pbckap}\ee

Using (\ref{1pbckap}) in (\ref{1psmatrelap}), we obtain

\bea
f^{10}_{a\alpha}(x-t-L)=S_{ab,\alpha\beta}(x-t)f^{10}_{b,\beta}(x-t).\label{1psmatrel2ap}
\eea
This is a matrix difference equation, which can be solved to obtain the amplitude $f^{10}_{a\alpha}(x-t)$. One can then use the equation (\ref{1pbckap}) to obtain the amplitude $f^{01}_{a\alpha}(x-t)$ and thereby obtain the one particle wavefunction. We will now consider the two particle case.

\section{Two particle wavefunction}

The wavefunction takes the form
\be\label{2pformkap}\ket{\Psi_2}=\sum_{ac\alpha}\int_{0}^L\int_0^Ldx_1 dx_2 \Psi^{\dagger}_a(x_1)\Psi^{\dagger}_c(x_2)\mathcal{A}F_{ac\alpha}(x_1,x_2,t)\ket{\alpha}, 
\ee

where $a,c$ denote the spin indices of the electrons and $\alpha$ denotes the spin of the impurity. $\mathcal{A}$ is the anti-symmetrizer under the exchange of $x_1\leftrightarrow x_2, a\leftrightarrow c$. Using the above equation (\ref{2pformkap}) in the Schrodinger equation (\ref{SEkap}), one obtains 

\begin{align} -i(\partial_t+\partial_{x_1}+\partial_{x_2})\mathcal{A}F_{ac\alpha}(x_1,x_2,t)&+J(t)\big(\delta(x_1)\:\vec{\sigma}_{ab}\cdot\vec{S}_{\alpha\beta}\:I_{cd}+\delta(x_2)\: \vec{\sigma}_{cd}\cdot\vec{S}_{\alpha\beta}\:I_{ab}\big)\mathcal{A}F_{bd\beta}(x_1,x_2,t)=0.\label{2pdiffkap}\end{align}

\subsubsection{A simple wavefunction for time dependent interaction strength $J(t)$}

Similar to the one particle case, ignoring the boundary conditions, we can start by looking at a simple solution in terms of the wave packets. One needs to distinguish between the amplitudes corresponding to different ordering of particles with respect to each other. We have

\begin{align}\nonumber
    F_{ac\alpha}(x_1,x_2,t)= &e^{-\frac{1}{2\sigma^2}\left(x_1-t+\tau_1\right)^2}e^{-\frac{1}{2\sigma^2}\left(x_2-t+\tau_2\right)^2} 
    \big(f^{102}_{ac\alpha}(-\tau_1,-\tau_2)\theta(-x_1)\theta(x_2)+f^{201}_{ac\alpha}(-\tau_1,-\tau_2)\theta(-x_2)\theta(x_1) \\\nonumber    
    &+f^{120}_{ac\alpha}(-\tau_1,-\tau_2)\theta(-x_1)\theta(-x_2)\theta(x_2-x_1)+f^{210}_{ac\alpha}(-\tau_1,-\tau_2)\theta(-x_1)\theta(-x_2)\theta(x_1-x_2)\\&
    +f^{012}_{ac\alpha}(-\tau_1,-\tau_2)\theta(x_1)\theta(x_2)\theta(x_2-x_1)+f^{021}_{ac\alpha}(-\tau_1,-\tau_2)\theta(x_1)\theta(x_2)\theta(x_1-x_2)\big)\label{2pap}.\end{align}

Using (\ref{2pap}) in the Schrodinger equation (\ref{2pdiffkap}), we obtain the following relations between the amplitudes

\bea \nonumber f^{201}(-\tau_1,-\tau_2)= S^{10}(-\tau_{1}) f^{210}(-\tau_1,-\tau_2),&  &f^{102}(-\tau_1,-\tau_2)= S^{20}(-\tau_{2}) f^{120}(-\tau_1,-\tau_2)\\ f^{012}(-\tau_1,-\tau_2)=S^{10}(-\tau_{1}) f^{102}(-\tau_1,-\tau_2),& &f^{021}(-\tau_1,-\tau_2)= S^{20}(-\tau_{2}) f^{201}(-\tau_1,-\tau_2).\label{2prel1wpap1}\eea

Here we have suppressed the spin indices for notational ease. In the above equations $S^{10}(-\tau_1), S^{20}(-tau_2)$ act in the spin spaces of electrons `1' and `2' and the impurity respectively, and are given by the same form as (\ref{smatimp1ap}). The consistency of the solution requires one to impose the following constraints

\bea f^{210}(-\tau_1,-\tau_2)= S^{12}(-\tau_{1},-\tau_{2}) f^{120}(-\tau_1,-\tau_2), &&f^{021}(-\tau_1,-\tau_2)=S^{12}(-\tau_{1},-\tau_{2})f^{012}(-\tau_1,-\tau_2).\label{2prel2wpap}
\eea 

Here $S^{12}(-\tau_1,-\tau_2)$ acts in the spin spaces of the electrons and is given by
\bea S^{12}(-\tau_{1},-\tau_{2})= \frac{i(g(-\tau_{1})-g(-\tau_{2}))I^{12}+P^{12}}{i(g(-\tau_{1})-g(-\tau_{2}))+1}.\label{2psmatwpap}
\eea

These S-matrices satisfy the Yang-Baxter equation
\bea  S^{12}(-\tau_{1},-\tau_{2})S^{10}(-\tau_{1})S^{20}(-\tau_{2})=S^{20}(-\tau_{2})S^{10}(-\tau_{1})S^{12}(-\tau_{1},-\tau_{2}).\label{wpybap}
\eea
Here we see that even though there exists no interaction term between the electrons in the Hamiltonian, the electrons are correlated with each other due to the non trivial S-matrix $S^{12}(-\tau_{1},-\tau_{2})$. 

\subsubsection{General wavefunction for time dependent interaction strength $J(t)$}

Similar to the one particle case, the most general two particle wave function can be obtained by creating a superposition of the wave packets for all values of the parameters $\tau_1$ and $\tau_2$ in (\ref{2pap}) as follows

\bea \int_{\tau_1}\int_{\tau_2} F_{ac\alpha}(x_1,x_2,t).
\eea

Considering the system with finite size and performing the above integrals, we obtain the following explicit form of the general two particle wave function

\begin{align}
\nonumber F_{ac\alpha}(x_1,x_2,t)=& f^{120}_{ac\alpha}(x_1-t,x_2-t)\:\theta(-x_2)\theta(-x_1)\theta(L-y+x_1)\theta(L-y+x_2)\theta(x_2-x_1)\\\nonumber +&f^{210}_{ac\alpha}(x_1-t,x_2-t)\:\theta(-x_2)\theta(-x_1)\theta(L-y+x_1)\theta(L-y+x_2)\theta(x_1-x_2)\\\nonumber+&f^{012}_{ac\alpha}(x_1-t,x_2-t)\:\theta(x_1)\theta(x_2)\theta(y-x_1)\theta(y-x_2)\theta(x_2-x_1)\\\nonumber +&f^{021}_{ac\alpha}(x_1-t,x_2-t)\:\theta(x_1)\theta(x_2)\theta(y-x_1)\theta(y-x_2)\theta(x_1-x_2)\\\nonumber+&f^{102}_{ac\alpha}(x_1-t,x_2-t)\:\theta(-x_1)\theta(x_2)\theta(L-y+x_1)\theta(y-x_2)\\\nonumber +&f^{201}_{ac\alpha}(x_1-t,x_2-t)\:\theta(-x_2)\theta(x_1)\theta(L-y+x_2)\theta(y-x_1).\\\label{2pwfkap} &(\text{most general two particle wave function} )
\end{align}

 Note that (\ref{2prel1wpap1},\ref{2prel2wpap},\ref{2psmatwpap}) and (\ref{wpybap}) are relations between the amplitudes corresponding to the simple two particle wave function in terms of the wave packets. For clarity, below we reproduce the analogous relations between the amplitudes corresponding to the general two particle wave function (\ref{2pwfkap}). Using (\ref{2pwfkap}) in (\ref{2pdiffkap}), we obtain the following equations (supressing the spin indices)

\begin{align}\nonumber
f^{201}(y-t,x_2-t)=f^{120}(y-t-L,x_2-t),\\\nonumber f^{102}(x_1-t,y-t)=f^{210}(x_1-t,y-t-L),\\ \nonumber f^{021}(y-t,x_2-t)=f^{102}(y-t-L,x_2-t),\\f^{012}(x_1-t,y-t)=f^{201}(x_1-t,y-t-L).\label{2pbckap}
\end{align}

In addition, we obtain the following equations 

\begin{align}\nonumber
f^{201}(-t,x_2-t)=S^{10}(-t)f^{210}(-t,x_2-t),\\\nonumber f^{102}(x_1-t,-t)=S^{20}(-t)f^{120}(x_1-t,-t),\\\nonumber
f^{012}(-t,x_2-t)=S^{10}(-t)f^{102}(-t,x_2-t),\\ f^{021}(x_1-t,-t)=S^{20}(-t)f^{201}(x_1-t,-t).
\end{align}

Here the S-matrices $S^{j0}(-t)$ are exactly same as that in the one particle case. We have

\begin{align} \label{smatkap}&S^{j0}(-t)=e^{i\phi(-t)}\frac{ig(-t)I^{j0}_{ab\alpha\beta}+P^{j0}_{ab\alpha\beta}}{ig(-t)+1},\\
&g(-t)= \frac{1}{2J(t)}\left(1-\frac{3}{4}(J(t))^2\right),\\&e^{i\phi(-t)}=\frac{1+\left(i/2J(t)\right)\left(1-(3/4)J^2(t)\right)}{iJ(t)-\left(1+(3/4)J^2(t)\right)}.
\end{align}

Through simple change of variables the above equations lead to the following equations

\begin{align}\nonumber
f^{201}(x_1-t,x_2-t)=S^{10}(x_1-t)f^{210}(x_1-t,x_2-t),\\\nonumber f^{102}(x_1-t,x_2-t)=S^{20}(x_2-t)f^{120}(x_1-t,x_2-t),\\\nonumber
f^{012}(x_1-t,x_2-t)=S^{10}(x_1-t)f^{102}(x_1-t,x_2-t),\\ f^{021}(x_1-t,x_2-t)=S^{20}(x_2-t)f^{201}(x_1-t,x_2-t).\label{2psmatkap}
\end{align}

As mentioned in the maintext, the consistency of the wave function required us to differentiate between the amplitudes which differ in the ordering of the particles with respect to each other. These are $f^{120}(x_1-t,x_2-t)$ and $f^{210}(x_1-t,x_2-t)$ and similarly, $f^{012}(x_1-t,x_2-t)$ and $f^{021}(x_1-t,x_2-t)$. These pairs of amplitudes are not constrained by the Hamiltonian due to the relativistic dispersion. To preserve integrability, one needs to choose a specific electron-electron S-matrix that relates these amplitudes. It takes the following form

\begin{align}\nonumber
f^{210}(x_1-t,x_2-t) =S^{12}(x_1-t,x_2-t)f^{120}(x_1-t,x_2-t), \\f^{021}(x_1-t,x_2-t) =S^{12}(x_1-t,x_2-t)f^{012}(x_1-t,x_2-t),\label{eemateqsap}\end{align}

where \be S^{12}(x_1-t,x_2-t)=\frac{i(g(x_1-t)-g(x_2-t))I^{12}_{ab,\alpha\beta}+P^{12}_{ab,\alpha\beta}}{ig(x_1-t)-ig(x_2-t)+1}.\label{smateeap}\ee
The amplitudes $f^{210}(x_1-t,x_2-t)$ and $f^{021}(x_1-t,x_2-t)$ are `physical' when $x_1>x_2$, and `unphysical' for $x_1<x_2$. Similarly, the amplitudes $f^{120}(x_1-t,x_2-t)$ and $f^{012}(x_1-t,x_2-t)$ are physical when $x_2>x_1$ and unphysical for $x_2<x_1$. The relation (\ref{eemateqsap}) means that for $x_1<x_2$, the amplitudes $f^{120}(x_1-t,x_2-t)$, $f^{012}(x_1-t,x_2-t)$ which are physical, are related to the unphysical amplitudes $f^{210}(x_1-t,x_2-t)$, $f^{021}(x_1-t,x_2-t)$ respectively when multiplied by the operator $S^{12}(x_1-t,x_2-t)$, which is the particle-particle S-matrix. Equivalently, for for $x_2<x_1$, the amplitudes $f^{210}(x_1-t,x_2-t)$, $f^{021}(x_1-t,x_2-t)$, which are physical, are related to the unphysical amplitudes $f^{120}(x_1-t,x_2-t)$, $f^{012}(x_1-t,x_2-t)$ respectively when multiplied by the operator $\left(S^{12}(x_1-t,x_2-t)\right)^{-1}$. Here $\left(S^{12}(x_1-t,x_2-t)\right)^{-1}$ is the inverse of $S^{12}(x_1-t,x_2-t)$. The electron-electron S-matrix (\ref{smateeap}) and the electron-impurity S-matrices (\ref{2psmatkap}) satisfy the Yang-Baxter algebra

\be S^{20}(x_2-t)S^{10}(x_1-t)S^{12}(x_1-t,x_2-t)=S^{12}(x_1-t,x_2-t)S^{10}(x_1-t)S^{20}(x_2-t).\ee

 Using the first and the last equations of (\ref{2psmatkap}), we have

\begin{align}
f^{021}(x_1-t,x_2-t)=S^{20}(x_2-t)S^{10}(x_1-t)f^{210}(x_1-t,x_2-t).
\end{align}

Similarly, from the second and the third equations in (\ref{2psmatkap}), we have
\begin{align}
f^{012}(x_1-t,x_2-t)=S^{10}(x_1-t)S^{20}(x_2-t)f^{120}(x_1-t,x_2-t).
\end{align}

Consider the third equation in (\ref{2psmatkap}). Using the second equation in (\ref{eemateqsap}), we have
\begin{align}
f^{021}(x_1-t,x_2-t+L)=S^{12}(x_1-t,x_2-t+L)S^{10}(x_1-t)f^{102}(x_1-t,x_2-t+L).
\end{align}
Using the second equation of (\ref{2pbckap}) in the above equation, we have

\begin{align}
f^{021}(x_1-t,x_2-t+L)=S^{12}(x_1-t,x_2-t+L)S^{10}(t-x_1)f^{210}(x_1-t,x_2-t).
\end{align}

From simple variable changes in the third and second equations of (\ref{2pbckap}), we have 

\begin{align} f^{102}(x_1-t-L,x_2-t+L)=f^{021}(x_1-t,x_2-t+L),\\ f^{210}(x_1-t-L,x_2-t)=f^{102}(x_1-t-L,x_2-t+L).
\end{align}

Using the above three equations, we have

 \bea f^{210}(x_1-t-L,x_2-t)=S^{12}(x_1-t,x_2-t+L)S^{10}(x_1-t)f^{210}(x_1-t,x_2-t),\label{2pdiffeqk3ap}
\eea 

where the particle `1' is transported around the system once. Using the first equation of (\ref{eemateqsap}) in the second equation of (\ref{2psmatkap}), we obtain

\begin{align}f^{102}(x_1-t,x_2-t)=S^{20}(x_2-t)S^{12}(x_2-t,x_1-t)f^{210}(x_1-t,x_2-t).\end{align}
Using the second equation of (\ref{2pbckap}) in the above equation, we have

\begin{align}f^{210}(x_1-t,x_2-t-L)=S^{20}(x_2-t)S^{12}(x_2-t,x_1-t)f^{210}(x_1-t,x_2-t),\label{2pdiffeqk4ap}    \end{align}
where the particle `2' is transported around the system once. The operators in the equations (\ref{2pdiffeqk3ap}, \ref{2pdiffeqk4ap}) form a set of difference equations corresponding to the amplitude $f^{210}(x_1-t,x_2-t)$. The construction described above can be readily generalized to the case of $N$ particles, which is provided in the main text.

\section{A solution to the consistency conditions}  
In this section we obtain an example of the interaction strength $J(t)$ for which the system is integrable. As mentioned in the main text, for the system to be integrable, the interaction strength $J(t)$ is constrained by the following consistency conditions on the operator $Z_j$
\begin{align}
&Z_i(z_1,...,z_j-L,...,z_N) Z_j(z_1,...,z_N)=
Z_j(z_1,...,z_i-L,...,z_N) Z_i(z_1,...,z_N),\label{commutationsap}\end{align}
where $Z_j$ is the transport operator
\begin{align}
    Z_j(z_1,...,z_N)= S^{jj+1}(z_j,z_{j+1}+L)S^{jj+2}(z_j,z_{j+2}+L)...S^{jN}(z_j,z_N+L)\times\\
   \times e^{-i\phi(z_j)}S^{j0}(z_j)S^{j1}(z_j,z_1)...S^{jj-1}(z_j,z_{j-1}),
  \label{transfermatap}
\end{align}
which constrains the amplitude 
$A^{N...j...10}(z_1,...,z_N)$ in the wavefunction through the matrix difference equation
\begin{align}A^{N...j...10}(z_1,...z_j-L,...,z_N)=Z_j(z_1,...,z_N) \: A^{N...j...10}(z_1,...z_j,...,z_N).\label{diffeq1ap}\end{align}
Without loss of generality, consider the left side of (\ref{commutationsap}) for $i<j$
\begin{align}\nonumber
&Z_i(z_1,...,z_j-L,...,z_N) Z_j(z_1,...,z_N)= S^{ii+1}(z_i,z_{i+1}+L)...S^{ij}(z_i,z_j)\\
\nonumber &S^{ij+1}(z_i,z_{j+1}+L)...S^{iN}(z_i,z_N+L)S^{i1}(z_i,z_1)..S^{ii-1}(z_i,z_{i-1}) \\& 
S^{jj+1}(z_j,z_j+L)...S^{jN}(z_j,z_N+L)S^{j1}(z_j,z_1)...S^{ji}(z_j,z_i)...S^{jj-1}(z_j,z_{j-1}) e^{-i\phi(z_i)}e^{-i\phi(z_j)}. \label{opexap1}    \end{align}
It is convenient to express the above equation in terms of the XXX R-matrices, which take the following form
\begin{align} R^{ij}(x)=b(x)I^{ij}+c(x)P^{ij}, \;\; b(x)=\frac{ix}{ix+1},\;\; c(x)=\frac{1}{ix+1}.\label{rmatdef}
\end{align} 
The particle-impurity (\ref{smatk1ap}) and particle-particle S-matrices (\ref{smateeap}) are related to the R-matrix through the following relations
\begin{align}
  S^{i0}(x_i-t)&= e^{i\phi(x_i-t)} R^{i0}(g(x_i-t)), \\ S^{ij}(x_i-t,x_j-t)&=R^{ij}(g(x_i-t)-g(x_j-t)).\label{srmatrel}
\end{align}
Using the definition of the R-matrix (\ref{rmatdef}) and the expressions for the particle-particle (\ref{smateeap}) and particle-impurity S-matrices (\ref{smatk1ap}), the above expression (\ref{opexap1}) can be expressed in terms of the R-matrices. We have
\begin{align}\nonumber
&Z_i(z_1,...,z_j-L,...,z_N) Z_j(z_1,...,z_N)= R^{ii+1}(g(z_i)-g(z_{i+1}+L))...R^{ij}(g(z_i)-g(z_j))\\\nonumber &R^{ij+1}(g(z_i)-g(z_{j+1}+L))...R^{iN}(g(z_i)-g(z_N+L)) R^{i1}(g(z_i)-g(z_1))..R^{ii-1}(g(z_i)-g(z_{i-1)})\\&
R^{jj+1}(g(z_j)-g(z_j+L))...R^{jN}(g(z_j)-g(z_N+L))R^{j1}(g(z_j)-g(z_1))...R^{jj-1}(g(z_j)-g(z_{j-1}))\end{align}
The R matrices on the left side in $Z_j$ that do not act in the spin space of the $i^{th}$ particle can be moved to the left past the R matrices on the right side in $Z_i$ up until one reaches the R-matrix in $Z_i$ that acts in the spin space of $j+1^{th}$ particle (Note that $i<j$). We have
\begin{align}\nonumber
&Z_i(z_1,...,z_j-L,...,z_N) Z_j(z_1,...,z_N)= R^{ii+1}(g(z_i)-g(z_{i+1}+L))...R^{ij}(g(z_i)-g(z_j))\\\nonumber&R^{ij+1}(g(z_i)-g(z_{j+1}+L))R^{jj+1}(g(z_j)-g(z_{j+1}+L))...R^{iN}(g(z_i)-g(z_N+L))\\\nonumber & R^{jN}(g(z_j)-g(z_N+L))R^{i1}(g(z_i)-g(z_1))R^{j1}(g(z_j)-g(z_1))... R^{ii-1}(g(z_i)-g(z_{i-1}))\\&R^{ji-1}(g(z_j)-g(z_{i-1}))...R^{jj-1}(g(z_j)-g(z_{j-1}))
.\end{align}
In the above expression consider 
\be R^{ij}(g(z_i)-g(z_j))R^{ij+1}(g(z_i)-g(z_{j+1}+L))R^{jj+1}(g(z_j)-g(z_{j+1}+L)).
\ee
The Yang-Baxter algebra can be applied such that the above expression can be written as
\be R^{jj+1}(g(z_j)-g(z_{j+1}+L))
R^{ij+1}(g(z_i)-g(z_{j+1}+L))R^{ij}(g(z_i)-g(z_j)).\ee
This allows us to move  $R^{jj+1}(g(z_j)-g(z_{j+1}+L))$ to the left past all the operators in $Z_i$. We have
\begin{align}\nonumber
&Z_i(z_1,...,z_j-L,...,z_N) Z_j(z_1,...,z_N)=R^{jj+1}(g(z_j)-g(z_{j+1}+L))\\\nonumber&R^{ii+1}(g(z_i)-g(z_{i+1}+L))...R^{ij+1}(g(z_i)-g(z_{j+1}+L))R^{ij}(g(z_i)-g(z_j))\\\nonumber&R^{ij+2}(g(z_i)-g(z_{j+2}+L))R^{jj+2}(g(z_j)-g(z_{j+2}+L))
...\\\nonumber& R^{iN}(g(z_i)-g(z_N+L))R^{jN}(g(z_j)-g(z_N+L))R^{i1}(g(z_i)-g(z_1))R^{j1}(g(z_j)-g(z_1))...\\\nonumber& R^{ii-1}(g(z_i)-g(z_{i-1}))R^{ji-1}(g(z_j)-g(z_{i-1}))R^{ji}(g(z_j)-g(z_i))...R^{jj-1}(g(z_j)-g(z_{j-1})).
\end{align}
Just as before, we can use the Yang-Baxter relation 
\begin{align}\nonumber
&R^{ij}(g(z_i)-g(z_j))R^{ij+2}(g(z_i)-g(z_{j+2}+L))R^{jj+2}(g(z_j)-g(z_{j+2}+L))\\&=R^{jj+2}(g(z_j)-g(z_{j+2}+L))R^{ij+2}(g(z_i)-g(z_{j+2}+L))R^{ij}(g(z_i)-g(z_j)),
\end{align}
and move $R^{jj+2}(g(z_j)-g(z_{j+2}+L))$ to the left past all the operators in $Z_i$. We have
\begin{align}\nonumber
&Z_i(z_1,...,z_j-L,...,z_N) Z_j(z_1,...,z_N)=R^{jj+1}(g(z_j)-g(z_{j+1}+L))\\\nonumber&R^{jj+2}(g(z_j)-g(z_{j+2}+L))R^{ii+1}(g(z_i)-g(z_{i+1}+L))...R^{ij+1}(g(z_i)-g(z_{j+1}+L))\\\nonumber&R^{ij+2}(g(z_i)-g(z_{j+2}+L))R^{ij}(g(z_i)-g(z_j))R^{ij+3}(g(z_i)-g(z_{j+3}+L))\\\nonumber& R^{jj+3}(g(z_j)-g(z_{j+3}+L))
...R^{iN}(g(z_i)-g(z_N+L))R^{jN}(g(z_j)-g(z_N+L))\\\nonumber&R^{i1}(g(z_i)-g(z_1))R^{j1}(g(z_j)-g(z_1))...R^{ii-1}(g(z_i)-g(z_{i-1}))R^{ji-1}(g(z_j)-g(z_{i-1}))\\&R^{ji}(g(z_j)-g(z_i))...R^{jj-1}(g(z_j)-g(z_{j-1})).
\end{align}
By following the same procedure as above, we obtain
\begin{align}
\nonumber
&Z_i(z_1,...,z_j-L,...,z_N) Z_j(z_1,...,z_N)=R^{jj+1}(g(z_j)-g(z_{j+1}+L))\\\nonumber&R^{jj+2}(g(z_j)-g(z_{j+2}+L))...R^{jN}(g(z_j)-g(z_N+L))R^{j1}(g(z_j)-g(z_1))...\\\nonumber&R^{ji-1}(g(z_j)-g(z_{i-1}))R^{ii+1}(g(z_i)-g(z_{i+1}+L))...R^{ij-1}(g(z_i)-g(z_{j-1}+L))\\\nonumber& R^{ij+1}(g(z_i)-g(z_{j+1}+L)R^{ij+2}(g(z_i)-g(z_{i+2}+L))...R^{iN}(g(z_i)-g(z_N+L))\\\nonumber&R^{i1}(g(z_i)-g(z_1)) R^{i2}(g(z_i)-g(z_2))...R^{ii-1}(g(z_i)-g(z_{i-1}))R^{ij}(g(z_i)-g(z_j))\\&R^{ji}(g(z_j)-g(z_i))R^{ji+1}(g(z_j)-g(z_{i+1}))...R^{jj-1}(g(z_j)-g(z_{j-1})).
\end{align}
We can now use the identity 
\be R^{ij}(g(z_i)-g(z_j))R^{ji}(g(z_j)-g(z_i))=1,\ee
which allows us to move the set of operators $R^{ji+1}(g(z_j)-g(z_{i+1}))...R^{jj-1}(g(z_j)-g(z_{j-1}))$ to the left up to the point where one reaches the R-matrix that acts in the spin space of $i+1^{th}$ particle. We obtain 
\begin{align}
\nonumber
&Z_i(z_1,...,z_j-L,...,z_N) Z_j(z_1,...,z_N)=R^{jj+1}(g(z_j)-g(z_{j+1}+L))\\\nonumber&R^{jj+2}(g(z_j)-g(z_{j+2}+L))...R^{jN}(g(z_j)-g(z_N+L))R^{j1}(g(z_j)-g(z_1))\\\nonumber&...R^{ji-1}(g(z_j)-g(z_{i-1}))R^{ii+1}(g(z_i)-g(z_{i+1}+L))R^{ji+1}(g(z_j)-g(z_{i+1}))\\\nonumber &
R^{ii+2}(g(z_i)-g(z_{i+2}+L))R^{ji+2}(g(z_j)-g(z_{i+2}))...R^{ij-1}(g(z_i)-g(z_{j-1}+L))\\\nonumber&R^{jj-1}(g(z_j)-g(z_{j-1}))R^{ij+1}(g(z_i)-g(z_{j+1}+L))...R^{iN}(g(z_i)-g(z_N)+L)\\&R^{i1}(g(z_i)-g(z_1))...R^{ii-1}(g(z_i)-g(z_{i-1})).
\end{align}
We now insert $1=R^{ji}(g(z_j+L)-g(z_i))R^{ij}(g(z_i)-g(z_j+L))$ between $ R^{jj-1}(g(z_j)-g(z_{j-1}))$ and $R^{ij+1}(g(z_i)-g(z_{j+1}+L))$. If the interaction strength $J(t)$ is such that 
\be g(z\pm L)=g(z)\pm\kappa, \label{solutionconsistencyap}\ee where $\kappa$ is a constant, then we can use the Yang-Baxter relation \begin{align}\nonumber  &R^{ij-1}(g(z_i)-g(z_{j-1})-\kappa))R^{jj-1}(g(z_j)-g(z_{j-1}))R^{ji}(g(z_j)+\kappa-g(z_i))\\&=R^{ji}(g(z_j)+\kappa-g(z_i))R^{jj-1}(g(z_j)-g(z_{j-1}))R^{ij-1}(g(z_i)-g(z_{j-1})-\kappa))
\end{align}
and obtain
\begin{align}
\nonumber
&Z_i(z_1,...,z_j-L,...,z_N) Z_j(z_1,...,z_N)=R^{jj+1}(g(z_j)-g(z_{j+1}+L))\\\nonumber&R^{jj+2}(g(z_j)-g(z_{j+2}+L))...R^{jN}(g(z_j)-g(z_N+L))R^{j1}(g(z_j)-g(z_1))\\\nonumber&...R^{ji-1}(g(z_j)-g(z_{i-1}))R^{ii+1}(g(z_i)-g(z_{i+1}+L))R^{ji+1}(g(z_j)-g(z_{i+1}))
\\\nonumber&R^{ii+2}(g(z_i)-g(z_{i+2}+L))R^{ji+2}(g(z_j)-g(z_{i+2}))...R^{ji}(g(z_j+L)-g(z_i))\\\nonumber&R^{jj-1}(g(z_j)-g(z_{j-1}))R^{ij-1}(g(z_i)-g(z_{j-1}+L))R^{ij}(g(z_i)-g(z_j+L))\\&...R^{iN}(g(z_i)-g(z_N)+L)R^{i1}(g(z_i)-g(z_1))...R^{ii-1}(g(z_i)-g(z_{i-1})).
\end{align}
By using (\ref{solutionconsistencyap}), and the Yang-Baxter relations, we can move $R^{ji}(g(z_j+L)-g(z_i))$ to the left till the point where one reaches the R-matrix that acts in the spin spaces of $j^{th}$ and $i-1^{th}$ particles. In addition, the R-matrices corresponding to $Z_i$ that were shuffled when using the Yang-Baxter relation can also be moved to the right. We obtain
\begin{align}
\nonumber
&Z_i(z_1,...,z_j-L,...,z_N) Z_j(z_1,...,z_N)=R^{jj+1}(g(z_j)-g(z_{j+1}+L))\\\nonumber&R^{jj+2}(g(z_j)-g(z_{j+2}+L))...R^{jN}(g(z_j)-g(z_N+L))R^{j1}(g(z_j)-g(z_1))...\\\nonumber&R^{ji-1}(g(z_j)-g(z_{i-1}))R^{ji}(g(z_j+L)-g(z_i)) R^{ji+1}(g(z_j)-g(z_{i+1}))...\\\nonumber&R^{jj-1}(g(z_j)-g(z_{j-1}))R^{ii+1}(g(z_i)-g(z_{i+1}+L))R^{ii+2}(g(z_i)-g(z_{i+2}+L))\\&...R^{iN}(g(z_i)-g(z_N+L))R^{i1}(g(z_i)-g(z_1))...R^{ii-1}(g(z_i)-g(z_{i-1})).
\end{align}
The right side of the above equation is precisely $Z_j(z_1,...,z_i-L,...,z_N)Z_i(z_1,...,z_N)$. Hence, we see that the consistency conditions are satisfied. In the above, we have used the constraint (\ref{solutionconsistencyap}). Therefore, we have shown that if the interaction strength $J(t)$ is such that (\ref{solutionconsistencyap}) is satisfied, then the system is integrable. Note that for $J(t)$ satisfying (\ref{solutionconsistencyap}), the matrix difference equation (\ref{diffeq1ap}) turns into quantum Knizhnik-Zamolodchikov equation. An example of the interaction strength $J(t)$ which satisfies (\ref{solutionconsistencyap}) is presented in the main text.

\end{widetext}

\end{document}